%% file: main.tex
\pdfoutput=1
\documentclass[tightenlines, twocolumn, notitlepage, nofootinbib, superscriptaddress]{revtex4-2}

\usepackage[dvipsnames]{xcolor}
\usepackage{amsmath}
\usepackage{amssymb}
\usepackage{bm}
\usepackage{bibunits}
\usepackage{changes}
\usepackage{enumerate}
\usepackage{floatrow}
\usepackage{hyperref}
\usepackage[symbol]{footmisc}
\usepackage{graphicx}
\usepackage{units}
\usepackage{xspace}
\usepackage{booktabs}
\usepackage{listings}
\usepackage[frozencache=true,cachedir=minted-cache]{minted} 
\usepackage{multirow}
\usepackage{gensymb}
\usepackage{algorithm2e}
\usepackage[utf8x]{inputenc}
\usepackage[strings]{underscore}
\usepackage[english]{babel}
\usepackage{colortbl}
\RestyleAlgo{ruled}
\usepackage{caption}
\usepackage{subcaption}

\usepackage{soul}

\DeclareFloatFont{small}{\scriptsize}
\newfloatcommand{capbtabbox}{table}[][\FBwidth]
\usepackage{blindtext}

\graphicspath{{./}, {figures/}}
\setcitestyle{super}

\newcommand{\dionysus}{\textsc{D{\footnotesize IONYSUS}}\xspace}
\newcommand{\rs}{$R^2$}
\newcommand{\yh}{\hat{y}}
\newcommand{\indicator}{1\!\!1}

\newcommand{\done}[0]{\cellcolor[gray]{0.8}}

\newcommand{\mycaption}[2]{\caption{\textbf{#1}. #2}}
\DeclareMathOperator*{\argmax}{arg\,max}
\DeclareMathOperator*{\argmin}{arg\,min}

\makeatletter
\renewcommand*{\p@subsection}{\thesection.}
\makeatother

\usepackage{array}
\usepackage{booktabs}
\SetKwComment{Comment}{/* }{ */}

\makeatletter
\renewcommand*{\p@subsection}{\thesection.}
\makeatother

\begin{document}

\begin{bibunit}[ieeetr] 

	\title{\large{Calibration and generalizability of probabilistic models on low-data chemical datasets with \dionysus }}
	

	\date{\today}
	
	\author{Gary Tom}
	\affiliation{Chemical Physics Theory Group, Department of Chemistry, University of Toronto, Toronto, ON, Canada}
	\affiliation{Department of Computer Science, University of Toronto, Toronto, ON, Canada}
	
	\author{Riley J. Hickman}
	\affiliation{Chemical Physics Theory Group, Department of Chemistry, University of Toronto, Toronto, ON, Canada}
	\affiliation{Department of Computer Science, University of Toronto, Toronto, ON, Canada}
	\affiliation{Vector Institute for Artificial Intelligence, Toronto, ON, Canada}

	\author{Aniket Zinzuwadia}
	\affiliation{Harvard Medical School, Harvard University, Boston, MA, USA}
	
	\author{Afshan Mohajeri}
	\affiliation{Department of Chemistry, Shiraz University, Shiraz, Iran}
	
	\author{Benjamin Sanchez-Lengeling}
	\affiliation{Google Research, Brain Team}
	
	\author{Al\'an Aspuru-Guzik}
	\email{alan@aspuru.com}
	\affiliation{Chemical Physics Theory Group, Department of Chemistry, University of Toronto, Toronto, ON, Canada}
	\affiliation{Department of Computer Science, University of Toronto, Toronto, ON, Canada}
	\affiliation{Vector Institute for Artificial Intelligence, Toronto, ON, Canada}
	\affiliation{Department of Chemical Engineering \& Applied Chemistry, University of Toronto, Toronto, ON, Canada}
	\affiliation{Department of Materials Science \& Engineering, University of Toronto, Toronto, ON, Canada}
	\affiliation{Lebovic Fellow, Canadian Institute for Advanced Research, Toronto, ON, Canada}

\input{abstract.tex}

	\maketitle

        \input{introduction.tex}
	
        \input{related_works.tex}

        \input{methods.tex}

        \input{experiments.tex}



\input{conclusion.tex}



	\section*{Data availability}
	
	The data and code used to produce the experiments in this work will be made available on the following GitHub repository \href{https://github.com/aspuru-guzik-group/dionysus}{https://github.com/aspuru-guzik-group/dionysus} under an MIT license.
	

	\section*{Acknowledgments}

    %
    G.T. acknowledges the support of Natural Sciences and Engineering Research Council of Canada (NSERC) through the Postgraduate Scholarships-Doctoral Program  (PSGD3-559078-2021).
    R.J.H. gratefully acknowledges NSERC for provision of the Postgraduate Scholarships-Doctoral Program (PGSD3-534584-2019), as well as support from the Vector Institute.
    A.A.-G. acknowledges support from the Canada 150 Research Chairs program and CIFAR, as well as the generous support of Anders G. Fr\"oseth.  
    Computations reported in this work were performed on the computing clusters of the Vector Institute and on the Niagara supercomputer at the SciNet HPC Consortium.~\cite{niagara1,niagara2} Resources used in preparing this research were provided, in part, by the Province of Ontario, the Government of Canada through CIFAR, and companies sponsoring the Vector Institute. SciNet is funded by the Canada Foundation for Innovation, the Government of Ontario, Ontario Research Fund - Research Excellence, and by the University of Toronto.

	\section*{Conflicts of interest}
	The authors declare no competing interests.


    \phantomsection\addcontentsline{toc}{section}{\refname}\putbib[main]
    
\end{bibunit}

\clearpage
\newpage

\begin{bibunit}
	
	\setcounter{page}{1}
	\setcounter{section}{0}
	\setcounter{subsection}{0}
	\setcounter{figure}{0} 
	
        \renewcommand{\thesection}{S.\arabic{section}}
	\renewcommand{\thefigure}{S\arabic{figure}}
	\renewcommand{\thetable}{S\arabic{table}}
 
	\onecolumngrid
	\subsection*{\normalsize{Supplementary Information}{\\}{\vspace{6pt}}
	 			\large{Calibration and generalizability of probabilistic models on low-data chemical datasets}{\\}{\vspace{6pt}}
			        \normalsize{\normalfont{Gary Tom,$^{1,2}$ Riley J. Hickman,$^{1,2,3}$ Aniket Zinzuwadia,$^{4}$ Afshan Mohajeri,$^{5}$ Benjamin Sanchez-Lengeling,$^{6}$ Al\'{a}n Aspuru-Guzik$^{1,2,3,7,8,9,*}$}}
			        {\\}{\vspace{6pt}}
			        \small{\normalfont{$^1$\textit{Chemical Physics Theory Group, Department of Chemistry, University of Toronto, Toronto, ON, Canada}}}{\\}
			        \small{\normalfont{$^2$\textit{Department of Computer Science, University of Toronto, Toronto, ON, Canada}}}{\\}
			        \small{\normalfont{$^3$\textit{Vector Institute for Artificial Intelligence, Toronto, ON, Canada}}}{\\}
           \small{\normalfont{$^4$\textit{Harvard Medical School, Harvard University, Boston, MA, USA}}}{\\}
           \small{\normalfont{$^5$\textit{Department of Chemistry, Shiraz University, Shiraz, Iran}}}{\\}
			        \small{\normalfont{$^6$\textit{Google Research, Brain Team}}}{\\}
			        \small{\normalfont{$^7$\textit{Department of Chemical Engineering \& Applied Chemistry, University of Toronto, Toronto, ON, Canada}}}{\\}
			        \small{\normalfont{$^8$\textit{Department of Materials Science \& Engineering, University of Toronto, Toronto, ON, Canada}}}{\\}
			        \small{\normalfont{$^9$\textit{Lebovic Fellow, Canadian Institute for Advanced Research, Toronto, ON, Canada}}}{\\}
			        \small{\normalfont{$^*$\textit{alan@aspuru.com}}}
			    }
	{\vspace{18pt}}

\input{si}

	\putbib[supp]
\end{bibunit}

\end{document}

%% file: abstract.tex
\begin{abstract}


Deep learning models that leverage large datasets are often the state of the art for modelling molecular properties. When the datasets are smaller ($< 2000$ molecules), it is not clear that deep learning approaches are the right modelling tool. In this work we perform an extensive study of the calibration and generalizability of probabilistic machine learning models on small chemical datasets. Using different molecular representations and models, we analyse the quality of their predictions and uncertainties in a variety of tasks (binary, regression) and datasets. We also introduce two simulated experiments that evaluate their performance: (1) Bayesian optimization guided molecular design, (2) inference on out-of-distribution data via ablated cluster splits. We offer practical insights into model and feature choice for modelling small chemical datasets, a common scenario in new chemical experiments. We have packaged our analysis into the \dionysus repository, which is open sourced to aid in reproducibility and extension to new datasets.


\end{abstract}

%% file: introduction.tex
\section{Introduction} \label{sec:introduction}

The design and discovery of molecular materials routinely enables technologies which have crucial societal consequences. Given a library of compounds, prediction of molecular functionality from its structure enables ranking and selection of promising candidates prior to experimental validation or other screening filters. Therefore, building accurate quantitative structure-activity relationship models (QSAR) is key to accelerated chemical design and efficient experimental decision-making.~\cite{muratov_qsar_2020} Models that leverage statistical patterns in data are now often the state of the art on such tasks. Specifically, data science and machine learning (ML) have played critical roles in modern science in general,~\cite{hey_fourth_2009} enabling the utilization of data at unprecedented scales. Deep learning (DL) models are able to extract statistical patterns in dataset features and give accurate QSAR predictions and classifications.~\cite{walters_applications_2021} When compared to traditional \textit{ab initio} techniques, such as density functional theory (DFT), ML models are less computationally demanding, and can learn statistical patterns directly from experimental data. However, the quality of such models is determined by the quality of the original datasets they are trained on, and thus the models are still affected by the cost of accurate data generation.

To date, many studies consider molecular property prediction tasks where training data is plentiful.~\cite{gilmer_neural_2017, busk2021calibrated} In real-world molecular design campaigns, particularly in the initial stages, only small molecular datasets ($\leq 2000$ data points) are available due to the expense associated with generating accurate property measurements (monetary, resource, labour, or ethical limitations). Despite the practical importance of this regime, molecular property prediction using ML with limited data instances has been relatively under-explored, and remains a challenging task, especially for deep learning models which often require large amounts of training instances due to large number of model parameters.

In the low-data setting, understanding a ML model’s performance is important since predictions inform decisions about further research directions, or, in a sequential learning setting, promote molecules to be subject to property measurement. In particular, we place emphasis on 1) the generalizability, the ability of a model to predict accurately on new chemical data, and 2) uncertainty calibration, the ability of a model to estimate the confidence of its predictions. 

\begin{figure}[b]
    \centering
    \includegraphics[width=\columnwidth]{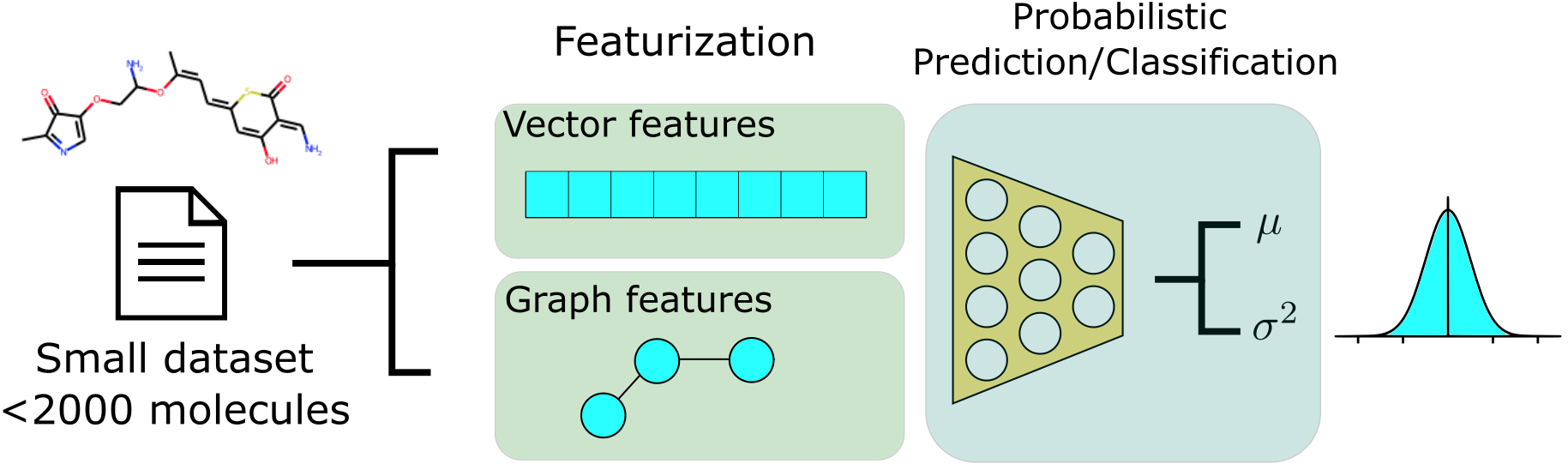}
    \mycaption{Schematic of the evaluation of probabilistic model on small molecular datasets}{We study the effects of different datasets, tasks, and featurizations. Probabilistic models output a prediction and an associated uncertainty.}
    \label{fig:big_picture}
\end{figure}

Adequate generalizability, the ability for a model to make accurate predictions on out-of-distribution (OOD) data, is paramount for many learning tasks, such as in the hit-to-lead and early lead optimization phases of drug discovery.\cite{nigam2021assigning, Graff:2021} After identification of a biological target (usually a protein or nucleic acid), initial molecular hits are optimized in an expensive and time-consuming make-design-test cycle. Using ML to predict molecular properties has indeed been shown to reduce the number of syntheses and measurements required.~\cite{schneider_rethinking_2020,sydow_advances_2019,varnek_machine_2012} Commonly, drug discovery project permit the synthesis and measurement of hundreds of candidate molecules due to constraints in expense, and typically involve functionalizations of a common molecular core or scaffold. Model generalization is therefore critical for the reuse of QSAR models for unstudied molecular scaffolds.~\cite{altae-tran_oneshot_2017,stanley2021fsmol}

Uncertainty calibration is the ability of a probabilistic model to produce accurate estimates of its confidence, and is also a crucial aspect of the molecular design process and high-risk decision making.~\cite{ovadia_trust_2019} Here, the goal is to learn a ML model that is not only accurate, but also furnishes its predictions with a notion of uncertainty. For instance, in a safety critical molecular property prediction scenario, e.g. the prediction of the severity of drug-induced liver injury,~\cite{williams_liver_2020,semenova_bayesian_2020} predictive uncertainty estimates can be an effective way of quantifying and communicating risk that can preserve time, resources, and human well-being. Additionally, strategies for sequential learning, such as Bayesian optimization~\cite{Mockus:1975,Mockus:1978,Mockus:2012} or active learning~\cite{settles_active_2009} commonly use uncertainties to construct utility functions, which determine how to promote molecules for property measurement based on their expected performance or informativeness. Previous studies have demonstrated that many state-of-the-art DL models are, although accurate, poorly calibrated.~\cite{guo_calibration_2017} Poorly calibrated predictions may have an adverse effect on decision-making.~\cite{silver_signal_2012}

	
We maintain that the topics of molecular property prediction in the low-data regime on one hand, and  uncertainty quantification and model generalizability on the other, are intimately related, as they are all commonly encountered in realistic molecular design and discovery campaigns. In the spirit of providing the community with a “handbook” on best practices thereof, we contribute \dionysus: a Python software package for facile evaluation of uncertainty quantification and generalizability of molecular property prediction models, accompanied by the current study, in which we showcase \dionysus by evaluating and analyzing these topics on several QSAR datasets.

The contributions of this work are as follows: 
\begin{itemize}
    \item We present a comprehensive study of the relationship between features and models in the low data regime across multiple axes: predictive performance, calibration of uncertainty, generalization and quality of uncertainty in optimization campaigns.
    \item We perform two experiments with associated metrics that can be conducted on generic regression and classification tasks: iterative molecule selection with Bayesian optimization and generalization on out-of-distribution (OOD) molecules. 
    \item We introduce a novel type of split to better benchmark predictive models on clusters of molecules.
    %
    \item This contribution describes our software which enables the extension of all analyses shown in this work to arbitrary molecular datasets. Most of the analysis is agnostic of ML model library and featurization strategy. 
    Code and experiments are packaged as \dionysus \href{https://github.com/aspuru-guzik-group/dionysus}{https://github.com/aspuru-guzik-group/dionysus}.
    \item We provide a “handbook” of practical recommendations for building and comparing models in the low-data regime.
\end{itemize}

%% file: related_works.tex
\section{Related work} \label{sec:related_work}

\subsection{Probabilistic models} \label{subsec:prob_models}

A variety of supervised learning models are available for representing predictive uncertainty. They can be broadly classified into two categories: those approaches derived from frequentist statistics and those based on Bayesian inference. 

Frequentist methods lack construction of a prior, and are instead concerned with the frequency of results over multiple trials. Ensemble methods are widely used examples of frequentist probabilistic machine learning.~\cite{dietterich2000ensemble} Ensemble-based methods generate uncertainty estimates based on the variance in the prediction of an ensemble of models that are trained on random subsets of data, as is the case of random forests (RF),~\cite{sheridan2012three, toplak2014assessment} or trained with randomly initialized parameters, as is often the case with weights of neural networks.~\cite{lakshminarayanan2017simple} For DL models, uncertainties can be estimated using Monte Carlo-dropout, in which the ensemble is created by randomly dropping out weights in a trained model at inference time.~\cite{gal2016dropout, cortes2019reliable} This approach is less computationally expensive, as it does not require training multiple neural networks with different weights.

Methods based on Bayesian inference seek to update a prior distribution, which summarizes pre-existing belief, in light of new observations. Commonly used Bayesian strategies for molecular property prediction in the low-data regime include Gaussian processes (GPs) \cite{rasmussen_gaussian_2006, hie2020leveraging, sanchez2019bayesian}, and Bayesian neural networks (BNNs).~\cite{blundell_weight_2015, zhang_bayesian_2019, ryu2019bayesian} GPs have more recently been combined with deep neural networks to produce more expressive models that naturally output uncertainties.~\cite{wilson2016deep, huang2015scalable} Several studies have highlighted the accuracy and calibration of such models on larger datasets.~\cite{liu_simple_2020, han2021reliable, busk2021calibrated}

\begin{figure*}[htb]
    \centering
\includegraphics[width=0.9\columnwidth]{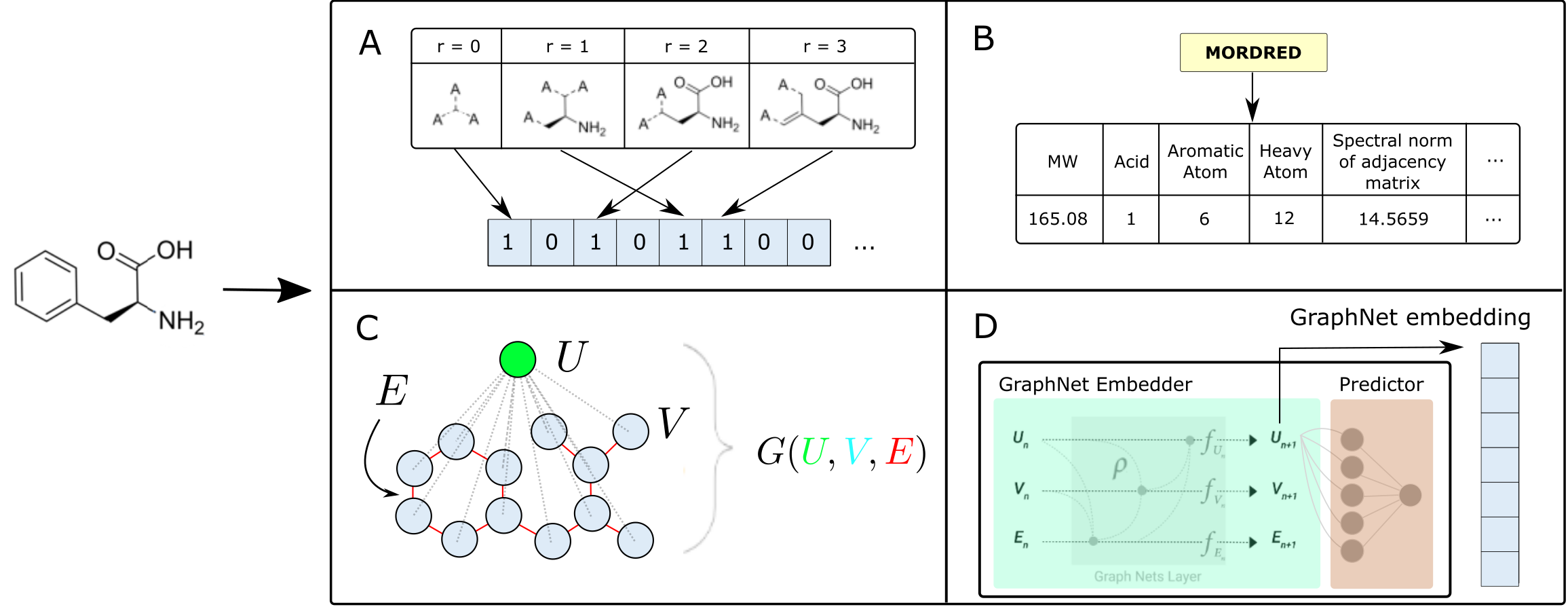}
    \mycaption{Schematic summary of molecular featurization methods}{All methods are available in \dionysus. A) Morgan fingerprints (MFP) are bit-vectors representing circular fragments of certain radii around heavy atoms.\cite{bajusz2017chemical} B) Mordred descriptors are physicochemical descriptors generated from the 2D graph. C) Graph tuple representations consist of the vertices (atoms) and edges (bonds) of a chemical graph, and the global node that is connected to all the atoms. D) Graph embeddings are extracted from the global node of a pretrained GraphNet GNN predictor.\cite{sanchez2021gentle}}
    \label{fig:molecular_features}
\end{figure*}

\subsection{Calibration and quantification of model uncertainties} \label{subsec:calib}

Despite the fact that many approaches exist to produce predictive uncertainties, they are not guaranteed to be calibrated. In fact, it is well known that many modern DL strategies are poorly calibrated, despite their accuracy.~\cite{guo_calibration_2017, nado2021uncertainty} For classification tasks, confidence calibration seeks to adjust probability estimates such that they reproduce the true correctness likelihood. Several calibration methods, such as isotonic regression \cite{zadrozny2002transforming}, Platt scaling,\cite{platt1999probabilistic} and temperature scaling \cite{niculescu2005predicting} can be applied as a learned post-processing step to any predictive model. Similar approaches can be extended confidence calibration to a regression task setting. 

Techniques have also been developed for ensuring ML models produce calibrated uncertainties through the use of regularization during training.~\cite{cui_calibrated_2020,soleimany_evidential_2021, hwang2020comprehensive} In one particular case, Soleimany et al.~\cite{soleimany_evidential_2021} leverage evidential deep learning ~\cite{sensoy_evidential_2018, amini_deep_2020} for molecular property prediction. While effective, such methods require careful choice of hyperparameters, as model confidence is sensitive to regularization strength. Multiple models must often be trained for each predictive task and molecular representation to determine the optimal evidential uncertainty hyperparameter(s).

Uncertainty quantification has since been studied for chemical prediction and classification tasks by numerous works.~\cite{moss_flowmo_2020,zhang_bayesian_2019} Hirschfield et al. studied and compared the performance of several neural network based uncertainty estimating techniques for regression on molecular properties.~\cite{hirschfeld_uncertainty_2020} Similarly, Hwang et al. employed graph neural networks (GNNs) for binary classification tasks on molecules.~\cite{hwang2020comprehensive} Similar issues with confidence calibration were observed, and corrections were applied through loss regularization.

\begin{table*}[htb] 
 \begin{center} 
\begin{ruledtabular}
\begin{tabular}{ccccc}
 Dataset name & Task type & Number of molecules & Heavy atom types  & Experimental value   \\ 
\hline
BioHL~\cite{mansouri2018opera}        & Regression             & 150   & C, S   & biodegradability half-life  \\
Freesolv~\cite{mobley_freesolv_2014}        & Regression             & 637   & C, N, O, F, P, S, Cl, Br, I   & free energy of solvation  \\
Delaney~\cite{delaney_esol_2004}            & Regression             & 1116  & C, N, O, F, P, S, Cl, Br, I   & log aqueous solubility \\
\hline
BACE~\cite{subramanian_computational_2016}  & Binary classification  & 1513  & C, N, O, F, S, Cl, Br, I   & binds to human BACE-1 protein \\
RBioDeg~\cite{mansouri2018opera}  & Binary classification  & 1608  &  C, N, O, F, Na, Si, P, S, Cl, Br, I  & readily biodegradable \\
BBBP~\cite{martins_bayesian_2012}           & Binary classification  & 1870  & B, C, N, O, F, P, S, Cl, Br, I   & blood-brain barrier permeability   \\
\end{tabular}
\end{ruledtabular}
\mycaption{Overview of the QSAR datasets considered in this work}{Both regression and binary classification tasks are explored. The datasets are all within the low-data regime ($<2000$ molecules). }
\label{tab:datasets}
\end{center}
\end{table*}

\subsection{Downstream applications of probabilistic models} \label{subsec:downstream}

Probabilistic ML models are the central component in real-world decision making. In the molecular design and discovery setting, they are commonly used in sequential learning frameworks, such as in high-throughput virtual screening, Bayesian optimization,\cite{Mockus:1975, Mockus:1978, Mockus:2012} and active learning.\cite{settles_active_2009} Common to these frameworks are a machine-learned surrogate model which approximates the true underlying structure-property relation, and a utility function which determines which molecules to subject to measurement based on their expected informativeness. Typically, utility functions balance exploitative and explorative sampling behavior by considering both the surrogate model's predictive mean and variance. 

Although such frameworks have been demonstrated in the context of molecular design and discovery, many applications have focused on tasks with large pools of available candidates. For example, Graff et al. report accelerated structure-based virtual screening large computational docking libraries ($> 10^8$ compounds) using scalable models trained using mean-variance estimation.~\cite{Graff:2021, graff_selffocusing_2022} Ryu et al. used Bayesian deep learning to screen the ChEMBL dataset~\cite{gaulton2017chembl} for active inhibitor molecules.~\cite{ryu2019bayesian} It was found that the Bayesian model returned active inhibitors at a significantly greater “hit rate” than did baseline strategies, suggesting that ML models with reliable uncertainty estimates execute more efficient screening campaigns. Studies considering smaller molecular datasets ($\leq 1000$) also exist. For instance, Zhang et al. used Bayesian semi-supervised graph convolutional neural networks to learn structure-bioactivity relationships with statistically principled uncertainty estimates.~\cite{zhang_bayesian_2019} The authors showed estimations of uncertainty in the low-data regime can drive an active learning cycle, obtaining low model error with limited queries for training set data. Despite the strong work reported in previous studies, the relationship between performance of an active learner and the calibration and generalizability of the surrogate model has been relatively underexplored, particularly in the low-data molecular setting.



%% file: methods.tex
\section{Methods} \label{sec:methods}


\begin{figure*}[htb]
    \centering
    \includegraphics[width=0.9\textwidth]{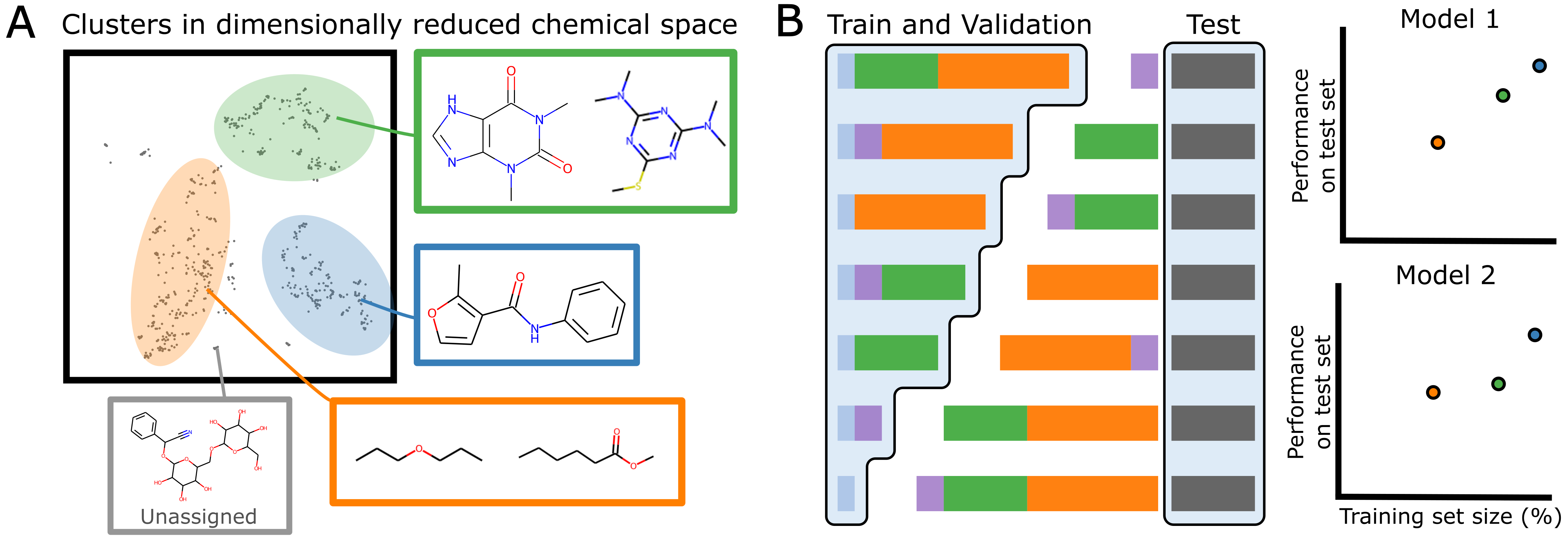}
    \mycaption{Visualization of molecular clusters splits}{A) Schematic of dimensionally reduced chemical space sorted by clusters of structurally similar molecules. B) Schematic for generating cluster splits of training/validation sets based on identified clusters. Performance and calibration is evaluated on the test set and plotted as function of available data.}
    \label{fig:cluster_splits}
\end{figure*}

\subsection{Molecular features} \label{subsec:molecular_feat}

Molecules must be represented in machine-readable format to enable computational property prediction. Several featurization methods are explored in \dionysus (Figure~\ref{fig:molecular_features}). All information is derived from the molecular graph, parsed from a SMILES string. The features used are categorized into 2 types: \textit{vector-valued} and \textit{graph-valued}. A $d$-dimensional vector-valued feature $\bm{x} \in \mathbb{R}^{d}$ comprise bit-vectors or physicochemical descriptors of a molecule, while graph-valued features are represented as a tuple $G = \left(  U, V, E \right)$. When referring abstractly to a molecular feature type, we use $X$ to represent either $\bm{x}$ or $G$ from herein.


Morgan fingerprints (MFPs) are generated by iterating over atomic centres and assigning bits according to neighboring structures up to a certain radius away.~\cite{rogers_fingerprints_2010} A hashing algorithm is then used to generate a high dimensional bit-vector unique to the molecule. For our experiments, we use $d=2048$ dimensional MFPs with radius 3, generated using the open-source cheminformatics package RDKit.\cite{landrum2022rdkit}

In addition to fingerprints, physicochemical molecular descriptors are often used for prediction of properties of molecules in cheminformatics techniques such as quantitative structure-activity/property relationship (QSAR/QSPR) models. We use the Mordred package to generate 1613 chemical descriptors from 2D molecular structures.~\cite{moriwaki_mordred_2018} 

The molecular graph can also be directly encoded in tuple graph representation, written as
%
$G = \left( U, V, E \right)$.~\cite{battaglia_relational_2018, sanchez2021gentle} The $d_u$-dimensional global attributes $U \in \mathbb{R}^{d_u}$ describe global properties of the molecule. $V$ is the set of node (atom) attributes $\{ \bm{v}_i\}_{i=1}^{N_v}$ for a molecule with $N_v$ atoms, where $\bm{v}_i \in \mathbb
{R}^{d_v}$. The set of edge (bond) features $E = \{ \left( \bm{e}_k, r_k, s_k  \right) \}_{k=1}^{N_e}$ comprise information about each of the $N_e$ bonds in the molecule. Here, $\bm{e}_k \in \mathbb{R}^{d_e}$ stores properties of the $k^{\text{th}}$ bond, while the indices $r_k$ and $s_k \in \{1, \cdots, N_v\}$ indicate the two vertices that the bond joins together. 
The atom and bond features used are listed in Table ~\ref{tab:node-edge-features}, while the global feature vector is zero-initialized. 

The tuples can be directly used as inputs for graph neural networks (GNN) predictor/classifier. From this representation, we also generate learned vector-based features, in which the graph-to-feature transformation is learned from the dataset. The graphs are passed through 3 GraphNet blocks and the resulting global vectors enter a prediction layer (Figure~\ref{fig:molecular_features}D). The global vectors from the trained network are the graph embeddings which are saved as vector-valued features for the various models.\cite{sanchez2019machine}



\subsection{Datasets and preprocessing} \label{subsec:datasets_preproc}

To evaluate model performance and calibration, we selected several datasets which contain experimentally determined properties for small organic molecules. The prediction task type, number of molecules, heavy atom types, and the chemical property measured are summarized in Table~\ref{tab:datasets}.
A dataset of $N$ molecules $\mathcal{D} = \{ \left( X_i, y \right) \}_{i=1}^{N}$ are comprised of pairs of molecular features $X$ and target properties $y \in \mathbb{R}$ for regression tasks and $y \in \{ 0, 1\}$ for binary classification tasks. 


\subsection{Models implemented} \label{subsec:models}


For each dataset and each featurization, we train and test five different models and evaluate the performance and uncertainty calibration of each: (1) NGBoost,~\cite{duan_ngboost_2019} (2) Gaussian process (GP),~\cite{rasmussen_gaussian_2006} (3) spectral-normalized GP (SNGP),~\cite{liu_simple_2020} (4) graph neural network GP (GNNGP),~\cite{han2021reliable} and (5) Bayesian neural networks (BNNs).~\cite{blundell_weight_2015} 
%

NGBoost is a random forest method that makes use of natural gradient boosting, similar to XGBoost\cite{chen2016xgboost}, to estimate the parameters of the conditional probability distribution of a certain observation given the input feature. An ensemble of 2000 decision trees with at most 3 layers comprise the ensemble, which will predict the parameters for a probability distribution; a Gaussian distribution for regression, and a Bernoulli distribution for binary classification. The ensemble is then fitted with the natural gradient to maximize the likelihood of the distribution for the given data. 

BNNs are probabilistic deep learning models that replace deterministic weight values of a neural network with weight distributions.~\cite{blundell_weight_2015} This allows the model to capture the epistemic uncertainty of the predictions due to limited training data. In our experiments, we use the local reparameterization estimator~\cite{kingma2015variational} to approximate the Gaussian prior distributions of 100 nodes of a single hidden layer. The output is passed through a rectified linear unit (ReLU) activation function and a final linear layer is used to generate a prediction.

A GP is a distance aware model in which predictions are made via transformed relative distances between data points, dictated by the kernel function. Unlike Bayesian deep learning, GPs allow for exact inference of the posterior predictive distribution. While exact inference is computational expensive for large datasets, exact GPs are effective in low-data regimes. We use the GPFlow package to implement the exact GP.~\cite{GPflow2017, GPflow2020multioutput} For the MFP features, we use a kernel based on the Tanimoto distance measure commonly used for high-dimensional binary vectors, which has been implemented in Moss \textit{et al.}~\cite{moss_flowmo_2020} The standard radial basis function (RBF) kernel was used for all other vector-valued features.

A SNGP is a deep kernel GP method, in which the kernel function is learned by training the model end-to-end.
The kernel is a multi-layer perceptron (MLP) with a spectral normalization procedure on the weights of each layer to inject distance awareness in the intermediate latent representations of the network.~\cite{liu_simple_2020} The features are then passed through a random features GP, which approximates a RBF kernel GP using a two-layer neural network.~\cite{rahimi_random_2007} 

A GNNGP is a graph-based model, trained end-to-end with the random features GP to combine the expressive power of graph representations with the probabilistic estimates of GPs. Like the GNN used to generate the graph embeddings, this model takes in graph tuples and has the same architecture as described above. The final predictive layer is replaced with the random features GP layers to produce predictive uncertainties.


\subsection{Evaluation metrics} \label{subsec:metrics}

\dionysus is designed in a modular way such that all predictions and uncertainties are saved, and metrics for the performance and calibration are calculated separately. Predictions and uncertainties from other models and datasets can be easily processed.

\subsubsection{Predictive metrics} \label{subsubsec:pred_metrics}

For regression tasks, previous works have utilized metrics such as root-mean-squared error or mean absolute error for measuring performance.~\cite{wu2018moleculenet} However, comparison of such metrics across target properties is often obfuscated by differences in magnitudes. As such, we use coefficient of determination \rs~ between a prediction and its ground truth. Values for \rs~ range from $-\infty$ to $1$. Models with \rs~ = 0 correspond to performance equal to the mean of the labels, while 1 corresponds to perfect prediction. Values can be lower than 0 since predictions can be infinitely worse.

Binary classification tasks are evaluated by the area under the receiver-operating curve (AUROC), which compares the true positive and true negative rates at different discrimination thresholds. An AUROC of $1.0$ indicates a perfect binary classifier, while a score of $0.5$ indicates random classifications. 

\subsubsection{Calibration metrics} \label{subsubsec:calib_metrics}

The calibration of a model is a measure of the correlation between the predicted uncertainty for a given input feature, and the error of the predicted value from the ground truth. For a well-calibrated model, the uncertainty associated with a poor prediction should be greater, and vice-versa. 


While there are many metrics that have been used, here we will use statistics generated from the reliability diagram, also known as the calibration diagram.~\cite{guo_calibration_2017} For regression tasks, the reliability diagram is given by the $C(q)$ score plotted as a function of quantile $q$, in which the Z-score statistic is used to compare the prediction and uncertainty with the ground truth.~\cite{moss_flowmo_2020} For a set of predictions $\hat{y}(X)$ with variances $\hat{\sigma}^2(X)$
\begin{equation}
    C(q) = \frac{1}{N} \sum_{i \in \mathcal{D}}  1\!\!1 \left( \left| \frac{\hat{y}(X_i) - y_i}{\hat{\sigma}(X_i)} \right| < \Phi^{-1} \left( \frac{1 + q}{2} \right) \right),
\end{equation}
where $\Phi^{-1}$ is the standard Gaussian inverse cumulative distribution, and $\indicator$ is the indicator function.

For the $q^{th}$ quantile, a well-calibrated model would have $q$ fraction of predictions Z-scores within the quantile, \emph{i.e.} $C(q) = q$. When $C(q) < q$, the model is overconfident, and when $C(q) > q$, the model is underconfident. The calibration metric obtained from the diagram is the absolute miscalibration area
\begin{equation}
    AMA(y, \yh, \hat{\sigma}) = \int_0^1  \Big| C(q, y, \hat{y}, \hat{\sigma}) - q \Big| ~ dq,
\end{equation}
which measures the absolute area between the model reliability and the perfect calibrated $C(q) = q$, with 0 area indicating a perfectly calibrated model.

For binary classification, the uncertainty of the model is given by the probability $p = \yh(X)$, or the mean of the Bernoulli distribution. The reliability diagram is given by the plot of the classification accuracy as a function of confidence $p$.~\cite{guo_calibration_2017} The predicted probabilities $p$ are binned into $M = 10$ uniform intervals $B_m$, where $m \in \{1 \cdots M\}$, and averaged for the confidence
\begin{equation}
    conf(B_m) = \frac{1}{|B_m|} \sum_{i\in B_m} \yh(X_i),
\end{equation}
while the accuracy is the fraction of correct classifications
\begin{equation}
    acc(B_m) = \frac{1}{|B_m|} \sum_{i \in B_m} \indicator(\yh(X_i) = y_i).
\end{equation}
Similar to the case of regression, we expect the accuracy to be equal to the given confidence, for example, at $p=0.5$, we would expect only half the predictions in the bin to be accurately classified.

The metric derived from the binary classification reliability diagram is known as the expected calibration error (ECE),~\cite{naeini2015obtaining} 
\begin{equation}
    ECE(y, \yh) = \sum_{m=1}^M \frac{|B_m|}{N} \Big| acc(B_m, y, \yh) - conf(B_m, y, \yh) \Big|,
\end{equation}
which is the average absolute difference between the accuracy and the confidence of the given bin, and is the discrete analog of the integral between the reliability curve and the perfectly calibrated model.

\begin{figure*}[htb]
    \centering
    \includegraphics[width=\columnwidth]{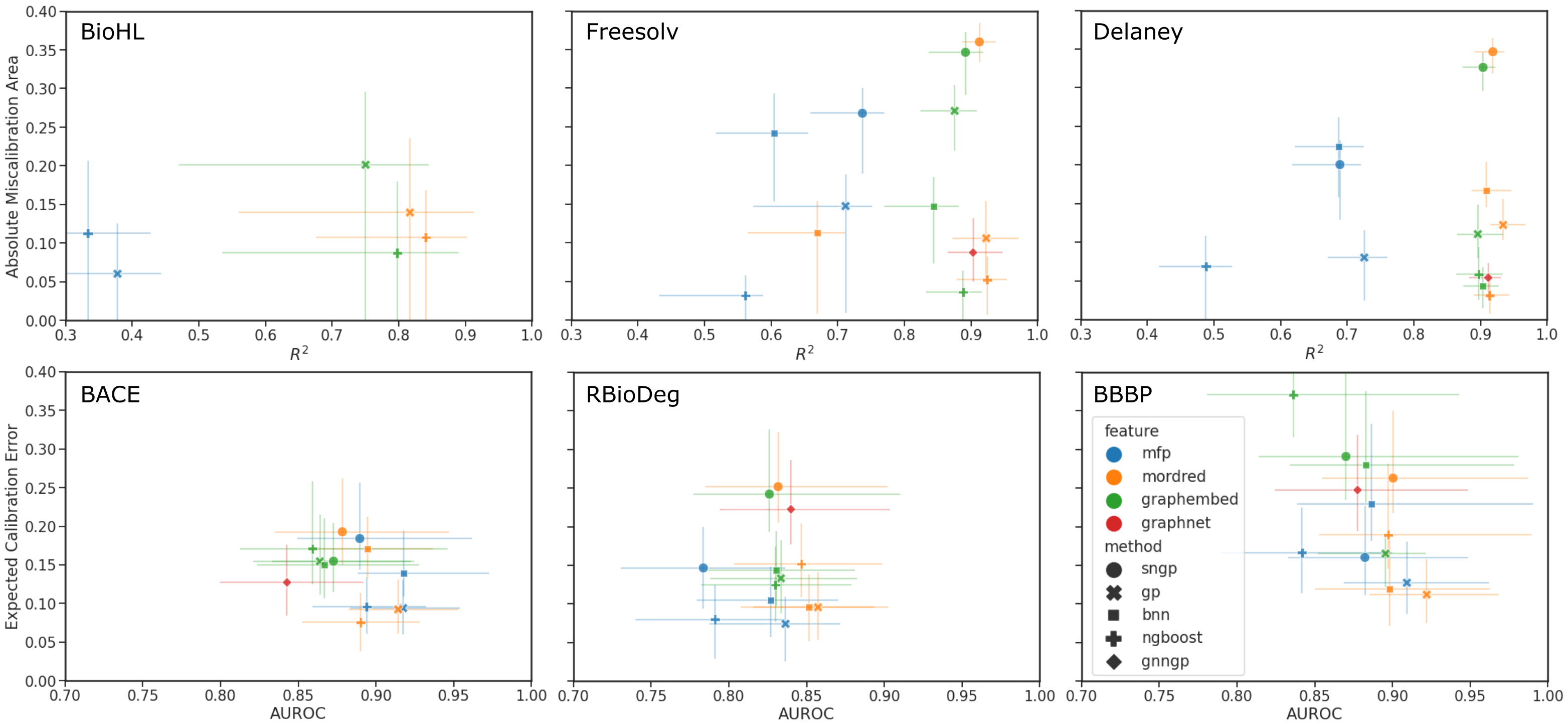}
    \mycaption{Plot of calibration error against model performance}{Results are for all models and compatible input features for regression and binary classification datasets.}
    \label{fig:regression_classification_results}
\end{figure*}

\subsection{Cluster splits} \label{subsec:cluster}

Datasets have often several structural motifs and we can identify them via a clustering algorithm. A cluster split separates a dataset into train and test via a cluster label. The test set will contain the structural motif and the training set will not. Performance on such splits can give 
an idea as to how well a model generalizes to new chemical classes. A cluster split can be viewed as a more general version of scaffold splitting which tends to involved a specific molecular core ~\cite{wu2018moleculenet}.

To build cluster splits, molecules are first assigned clusters based on the MFPs, which are decomposed into lower dimensional bit-vectors representing structural motifs using latent Dirichlet allocation.~\cite{hoffman2010online} 
The vectors are then appended to a bit representation of the dataset labels: for regression tasks, the values are binned into 10 discrete one-hot categories, and for binary classification, the classification label is used. This ensures that the individual clusters will not have extremely imbalanced labels. The joint labels are then further decomposed onto a 5-dimensional manifold using UMAP,~\cite{mcinnes_umap_2020} and the resulting vectors are clustered with the HDBScan algorithm.~\cite{campello2013density} The number of molecules in each cluster varies, and similar structures are indeed found within the same cluster (Figure~\ref{fig:cluster_splits}A). 

To evaluate the generalizability, a test set is separately generated by iterative stratified splitting 20\% of the dataset appended to the cluster labels, creating an unbiased test set across all the clusters.~\cite{szymanski2017network} The remaining molecules form various clusters which are partitioned into combinations of differently sized training sets (Figure~\ref{fig:cluster_splits}B). Validation set is a 15\% iterative stratified split of the training set.

%% file: experiments.tex
\section{Experiments and Results} \label{sec:experiments}

\input{supervised_learning}

\input{bayes_opt}

\input{generalization}

%% file: supervised_learning.tex
\subsection{Predictive performance and uncertainty calibration} \label{subsec:exp_supervised_learning}

In preparation for supervised learning experiments, regression datasets were randomly split into 70/10/20 percent training/validation/testing sets, while binary classification datasets were split using stratified splitting to ensure a similar proportion of classes in all three sets. Each model is trained with the described featurizations until an early stopping criteria is reached on the validation set to prevent overfitting to the training set. Finally, the predictions and uncertainties are made and saved for the testing set, and the models are evaluated by the aforementioned performance and calibration metrics. The $95\% $ confidence intervals were generated by bootstrapping from the test set results.




\begin{figure*}[htb]
    \centering
\includegraphics[width=0.8\columnwidth]{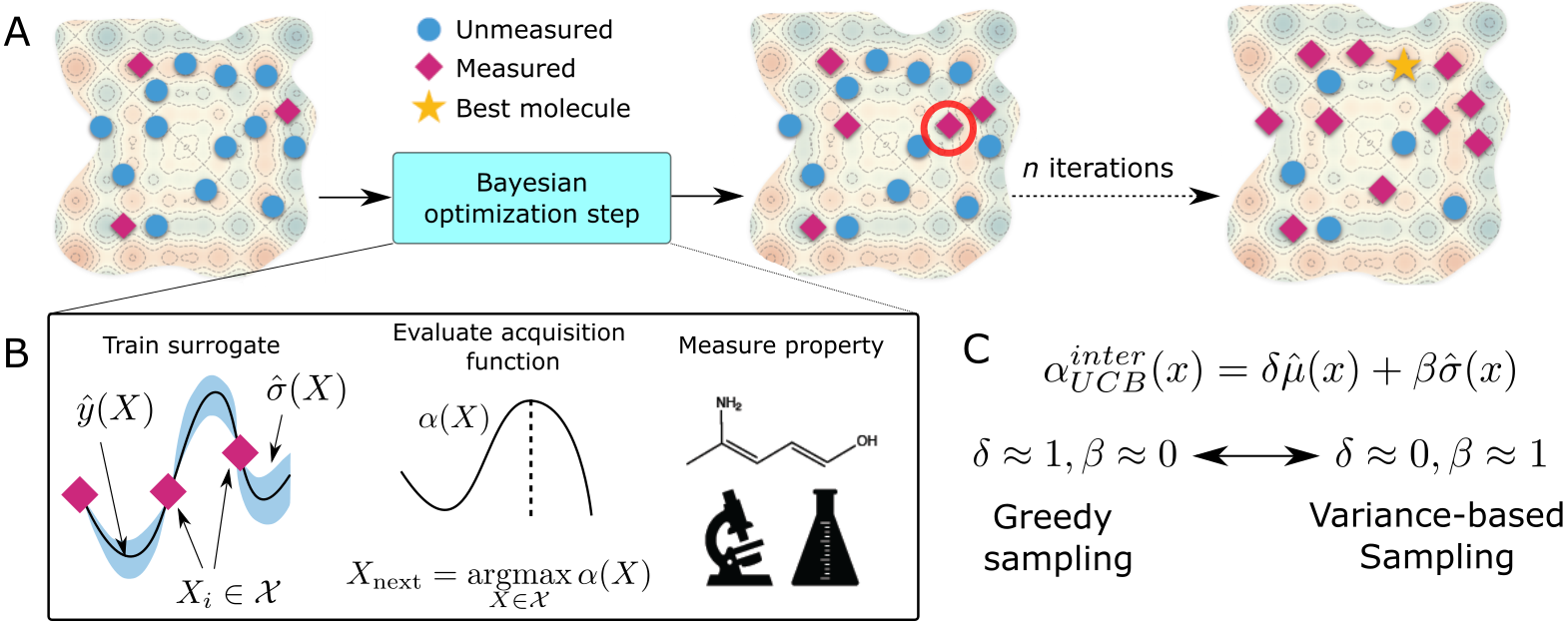}
    \mycaption{Bayesian optimization guided molecular design experiment}{A) BO pipeline with library of molecular candidates. Blue circles represent unmeasured molecular candidates, while red diamonds represent candidates for which a property measurement has transpired. The gold star indicates the structure with optimal parameters after termination of the optimization campaign.  B) Single BO step. C) Modified upper confidence bound acquisition function with interpolation parameters $\delta$ and $\beta$, where $\delta \equiv 1-\beta$.}
    \label{fig:bo_concept}
\end{figure*}

Plots comparing the calibration and performance metrics are shown in Figure~\ref{fig:regression_classification_results} for each of the datasets. The models and features with the best performance are found in the lower right of each plot, where the calibration error is minimized and the performance metric is maximized. The results are also tabulated in Tables~\ref{tab:regression_performance} for regression, and in Tables~\ref{tab:binary_performance} for classification.

In the regression data, we can observe much wider ranges in the performance metrics, particularly in the lower data regime of the BioHL and Freesolv datasets, with $R^2 < 0.3$ being truncated from the plot. The MFP feature has markedly lower $R^2$ scores and comparable AMA, with the Tanimoto kernel GPs performing the best. In the case of the BioHL dataset, all deep learning models (SNGP, BNN, GNNGP) struggled to compete with GPs and the NGBoost models trained on mordred descriptors and, surprisingly, graph embeddings, despite the small amount of data available. GNNGPs and BNNs on mordred and graph embeddings achieve competitive results in Freesolv and Delaney, likely due to the larger amount available training data. In all three regression datasets, the SNGP models achieve poor calibration, with high $R^2$ scores in Freesolv and Delaney.

In the binary classification data, the AUROC of all models and features are similar, likely due to the larger size of the datasets, and more data points representing the two binary classes in the discrete target than the continuous target, relative to the regression datasets. The error bars in the ECE score are much larger than those of the AMA in regression, since, in the low-data regime, there may be sparsely populated bins in the reliability diagram, and hence greater variability when bootstrapping. The best results are observed in GPs and NGBoost models trained on mordred and MFPs, possibly due to the importance of certain fragments represented by the MFP in the classification tasks represented here. Within the MFP results, we observe the best performance in the Tanimoto kernel GPs. Graph embeddings for all models gave higher calibration error. Among the deep learning models, SNGP and GNNGPs achieved good AUROC scores, but poorer calibration, while BNNs, when provided mordred descriptors, performed comparably to GPs and NGBoost models. We also observe an overall increase in classification miscalibration as the dataset size increases.




%% file: bayes_opt.tex
\subsection{Bayesian optimization} \label{subsec:bayes_opt}


\begin{figure}[b]
    \centering
    \includegraphics[width=\textwidth]{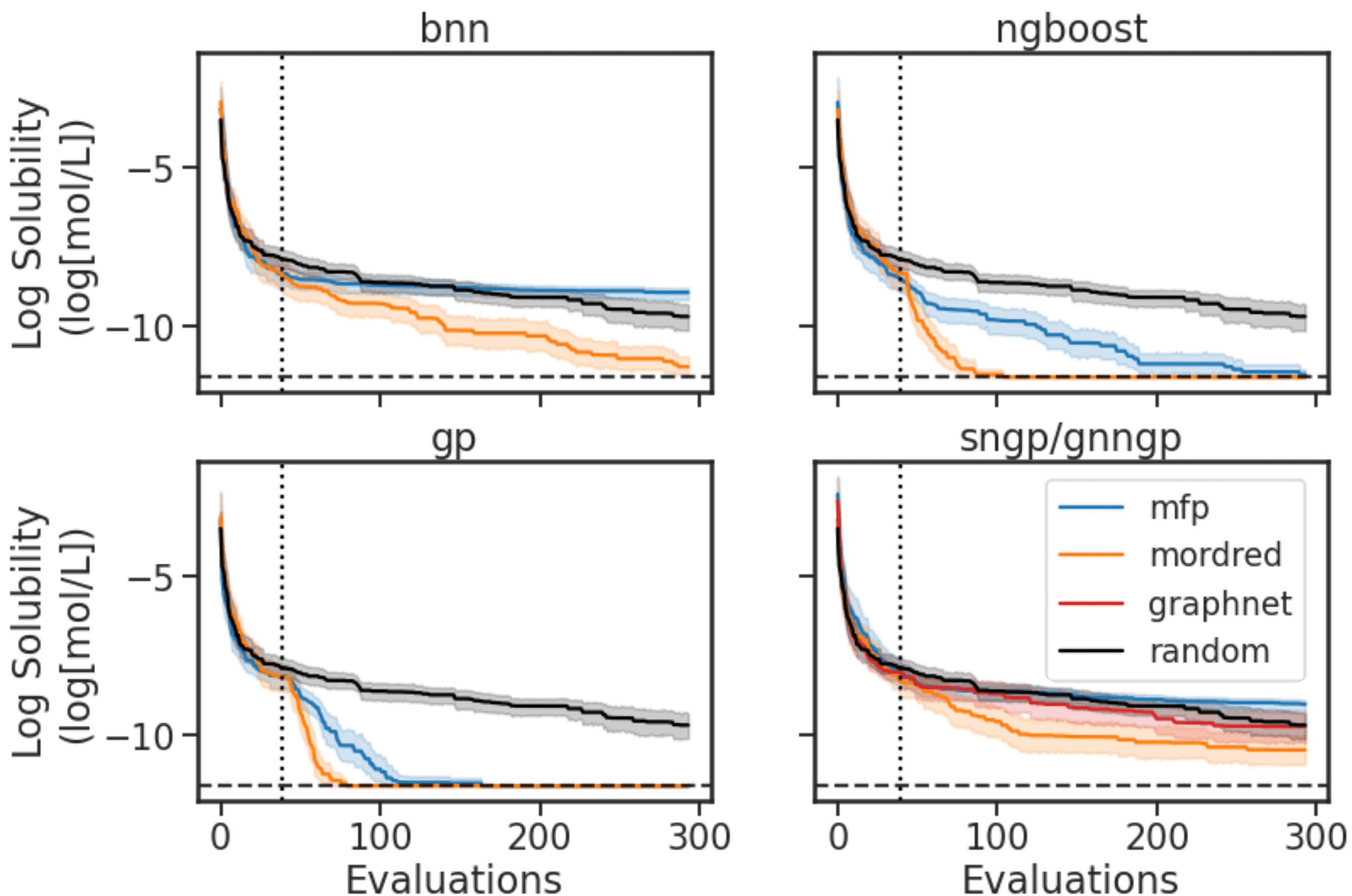}
    \mycaption{Optimization traces for the BO experiments on the Delaney dataset}{The goal to minimize the measured log solubility. Traces show average values over 30 independently seeded runs, and shaded areas show the $95\%$ confidence interval. Horizontal dashed lines indicate the optimal log solubility. The vertical dashed lines indicate the initial 5\% of the dataset.}
    \label{fig:delaney_bo_trace}
\end{figure}

Bayesian optimization (BO) is a global, model-based optimization strategy which consists of two main steps: 1) the inference of a probabilistic surrogate model to the unknown objective function based on all current measurements, and 2) the selection of new candidates for subsequent measurements using an acquisition function which balances the expected performance of each candidate and uncertainty of the surrogate model.
BO has been employed as a promising optimization framework across multiple disciplines,~\cite{Shahriari:2016} including automatic machine learning,~\cite{thornton_proceedings_2013,feurer2019auto,hutter2019auto} robotics,~\cite{calandra_bayesian_2016,berkenkamp_bayesian_2021} and experimental design.~\cite{vanlier_2012_experiment,foster_2019_experiment} More recently, BO has been employed to efficiently search through libraries of candidate molecules for those candidates which exhibit optimal properties.~\cite{hase_gryffin_2021,Graff:2021,shields_bayesian_2021}  Formally, for the minimization of a molecular property over candidate space $\mathcal{X}$, the optimization problem is
\begin{equation} \label{eq:opt_definition}
   	 X^* = \argmin_{X \in \mathcal{X}} f(X) \, ,
\end{equation}
where $f(\cdot)$ is some unknown black-box response function which in general is expensive to evaluate and potentially subject to noise (although we do not explicitly consider measurement noise here). We consider optimization over a domain $\mathcal{X}$ which consists of a finite set of $N$ molecular candidates defined \textit{a priori} to experimentation, i.e. $\mathcal{X} = \left\{ X_i \right\}_{i=1}^N$ (Figure~\ref{fig:bo_concept}). At each iteration, newly evaluated molecules are appended to a dataset of $K$ input-output pairs, $\mathcal{D} = \left\{ ( X_i, y_i) \right\}_{i=1}^K$, which is used to train the surrogate model.

The mean (prediction) and variance (uncertainty) of the model output are used to calculate the acquisition function. A plethora of acquisition functions have been proposed for BO. 
We consider the commonly used upper confidence bound (UCB)
\begin{align} \label{eq:conv_ucb}
    \alpha_{{UCB}} ( X ) = \hat{y}( X ) + \beta \hat{\sigma}(X) ,
\end{align}
which has a trade-off parameter $\beta$ that controls the contribution of the predicted variance in the acquisition function. This is set to 0.25 for the experiments. Molecule(s) recommended for measurement are those that maximize Eq.~\ref{eq:conv_ucb}, \textit{i.e.} $X_{\text{next}} = \argmax_{ X \in \mathcal{X}} \alpha_{\text{UCB}}( X )$. 

Representative results of simulated BO experiments using the Delaney dataset are shown in Figure \ref{fig:delaney_bo_trace}. Optimization traces for experiments on the remaining datasets are shown in \ref{sisubsec:bo_traces}. The algorithm aims to minimize the aqueous solubility, finding the molecule within the Delaney dataset that has the lowest water solubility. Traces represent the cumulative best log aqueous solubility value identified by each strategy, averaged over 30 independently seeded runs. For regression, the initial design dataset comprises $5\%$ of the dataset size and is randomly sampled. For binary classification, we start with $10\%$ to avoid only sampling molecules of the same class. For comparison, a random search was also performed. Note that the graph embeddings were not used due to the poor performance of GNN embedder at extremely low data regimes ($\sim 50$). 

In Figure \ref{fig:delaney_bo_trace}, we see the best performances with the Mordred descriptors, and the GP and NGBoost models. Similar to the results observed in the performance/calibration experiments, the MFP performs best with the Tanimoto kernel GP model. The deep learning models struggle with the sparse data: BNN and SNGP perform best with the Mordred descriptors, and perform no better than random search with MFPs. The GNNGP is unable to achieve better optimization with the graph inputs.

To succinctly summarize all experiments, the number of hit molecules are recorded for the optimization trace over the separate runs, and the results for the datasets of interest are shown in Table~\ref{tab:bo_regression_hits} and \ref{tab:bo_classification_hits}. In a regression task, a molecule is considered a hit if it is within the top 10\% of the dataset. In a classification task, as we are using a greedy strategy, a hit is a positive binary label. 

\begin{table}[t] 
 \begin{center} 
\begin{ruledtabular}
\begin{tabular}{cccc}
\textbf{BioHL} & MFP & Mordred & Graph   \\ 
\hline 
Random & $3.13 \pm 0.30$ & \done & \done \\
SNGP & $ 2.88 \pm 0.43 $ & $ 3.20 \pm 0.31 $ & \done  \\
GP & $ 3.43 \pm 0.25$  & $ 3.30 \pm 0.25$  &  \done  \\
BNN & $3.57 \pm 0.37$ & $3.40 \pm 0.24$ & \done   \\
NGBoost & $3.10 \pm 0.23$ & $3.47 \pm 0.20$ & \done    \\
GNNGP &\done &\done & $ 3.20 \pm 0.27$  \\
\hline
\textbf{Freesolv} & MFP & Mordred & Graph \\ 
\hline 
Random & $27.17 \pm 1.02$ & \done & \done \\
SNGP & $38.97 \pm 1.26$ & $35.57 \pm 1.84$ & \done  \\
GP & \bm{$49.23 \pm 0.56$}  & \bm{$49.50 \pm 0.51$}  & \done \\
BNN & $41.23 \pm 0.60$ & $40.73 \pm 0.91$ & \done \\
NGBoost & $47.17 \pm 0.66$ & \bm{$49.57 \pm 0.54$} & \done   \\
GNNGP &\done &\done & $24.90 \pm 1.37$  \\
\hline
\textbf{Delaney} & MFP & Mordred & Graph  \\ 
\hline
Random & $25.53 \pm 1.51$ & \done & \done \\
SNGP & $33.93 \pm 1.59$ & $31.40 \pm 2.03$ & \done  \\
GP & $76.67 \pm 1.13$  & \bm{$87.63 \pm 0.68$}  & \done \\
BNN & $35.83 \pm 1.66$ & $37.90 \pm 2.03$ & \done \\
NGBoost & $71.77 \pm 1.14$ & \bm{$88.17 \pm 0.63$} & \done   \\
GNNGP &\done &\done & $17.20 \pm 1.33$  \\
\end{tabular}
\end{ruledtabular}
\mycaption{Number of hits in Bayesian optimization of regression datasets}{The UCB acquisition function was used, with $\beta=0.25$. Statistics are gathered over all 30 runs, and the 95\% confidence interval is reported. The run starts with $5\%$ randomly sampled portion of the datasets.}
\label{tab:bo_regression_hits}
\end{center}
\end{table}

\begin{table}[h!] 
 \begin{center} 
\begin{ruledtabular}
\begin{tabular}{cccc}
\textbf{BACE} & MFP & Mordred & Graph   \\ 
\hline 
Random & $117.13 \pm 2.61$ & \done & \done \\
SNGP & $143.47 \pm 2.56$ & $130.73 \pm 3.22$ & \done  \\
GP & $200.07 \pm 1.85$ & \bm{$205.50 \pm 2.10$ } &  \done  \\
BNN & $146.80 \pm 3.04$ & $126.50 \pm 2.73$ & \done   \\
NGBoost & $194.87 \pm 1.78$ & $192.10 \pm 2.10$ & \done    \\
GNNGP &\done &\done & $122.27 \pm 2.43$  \\
\hline
\textbf{RBioDeg} & MFP & Mordred & Graph-based  \\ 
\hline 
Random & $106.77 \pm 2.87$ & \done & \done \\
SNGP & $114.17 \pm 2.34$ & $110.70 \pm 3.32$ & \done  \\
GP & \bm{$219.17 \pm 1.11$} & $214.67 \pm 1.67$  &  \done  \\
BNN & $111.97 \pm 2.20$ & $115.50 \pm 4.11$ & \done   \\
NGBoost & $198.87 \pm 2.37$ & $119.33 \pm 2.05$ & \done    \\
GNNGP &\done &\done & $126.17 \pm 2.60$  \\
\hline
\textbf{BBBP} & MFP & Mordred & Graph-based  \\ 
\hline
Random & $192.00 \pm 2.37$ & \done & \done \\
SNGP & $178.87 \pm 2.24$ & $184.3 \pm 3.93$ & \done  \\
GP & \bm{$243.80 \pm 0.73$} & $240.97 \pm 0.72$  &  \done  \\
BNN & $178.00 \pm 2.58$ & $189.90 \pm 2.90$ & \done   \\
NGBoost & $227.47 \pm 2.24$ & $217.83 \pm 2.92$ & \done    \\
GNNGP &\done &\done & $196.73 \pm 2.68$  \\
\end{tabular}
\end{ruledtabular}
\mycaption{Number of hits in Bayesian optimization of binary classification datasets}{A greedy strategy was used. Statistics are gathered over all 30 runs, starting from $10\%$ randomly sampled portion of datasets. The 95\% confidence interval is reported.}
\label{tab:bo_classification_hits}
\end{center}
\end{table}

In BioHL, due to the small data size, the random search is a relatively efficient search method, and is able to find the optimal molecule. Therefore, the number of hits are similar to those of the various surrogate models. In the Freesolv and Delaney datasets, all surrogate models optimize similar to or better than the random search, with the exception of GNNGP. The highest number of hits were achieved by NGBoost and GPs using mordred descriptors.

In the binary classification datasets, the highest number of hits are found by GP and NGBoost surrogate models trained on MFP and Mordred descriptors, with GPs performing slightly better, particularly in the RBioDeg and BBBP datasets. Again, the GPs perform better with MFP through the Tanimoto kernel. In general, the deep models perform similar to random search, indicating ineffective surrogate models. However, we consistently observe better optimization for all surrogate models using MFP for the BACE dataset, similar to the results of the performance/calibration experiments (Section \ref{subsec:exp_supervised_learning}).


\begin{figure}[h!]
    \centering
    \includegraphics[width=0.9\textwidth]{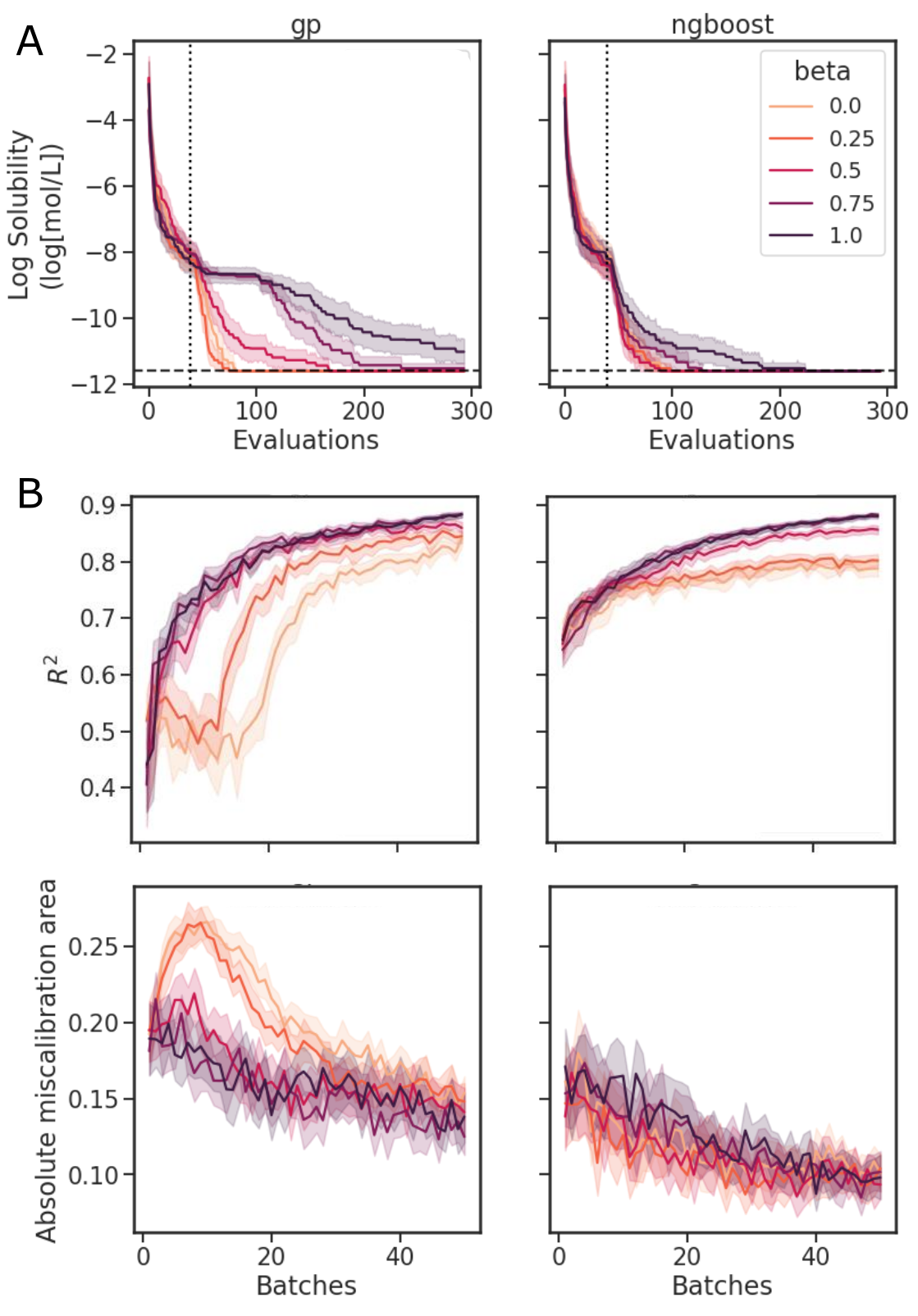}
    \mycaption{Results of $\beta$ parameter scan in the interpolated UCB acquisition function}{Metrics evaluated on test sets over the course of BO on Delaney dataset for GP and NGBoost models on Mordred descriptors. Shaded areas represent 95\% confidence intervals over 30 runs. A) The optimization traces. B) The performance and calibration metrics after each batch of measurements on a separate test set. }
    \label{fig:delaney_ucb_scan}
\end{figure}

To study the effects of the acquisition function on the surrogate model performance and calibration over the course of optimization, we use a modified UCB acquisition function,
\begin{equation}
    \alpha_{UCB}^{inter}(X) = \delta \hat{y}(X) + \beta \hat{\sigma}(X)
\end{equation}
which allows for interpolation among selection strategies that emphasize the predictive mean value and those that emphasize the predictive uncertainty. The parameter $\beta$ is scanned between values of 0 and 1, and $\delta \equiv 1 - \beta$. It is important to note that we normalize the values of both $\hat{y}$ and $\hat{\sigma}$ across the entire molecular candidate pool such that their values can be considered on equal footing. As $\beta$ approaches 0, greater weight is placed on the predictive mean, and the sampling behaviour should resemble that of a ``greedy" strategy. As $\beta$ approaches 1, greater emphasis is placed on the predictive uncertainty (Figure~\ref{fig:bo_concept}C).

\begin{figure*}[hbt]
    \centering
    \includegraphics[width=0.8\textwidth]{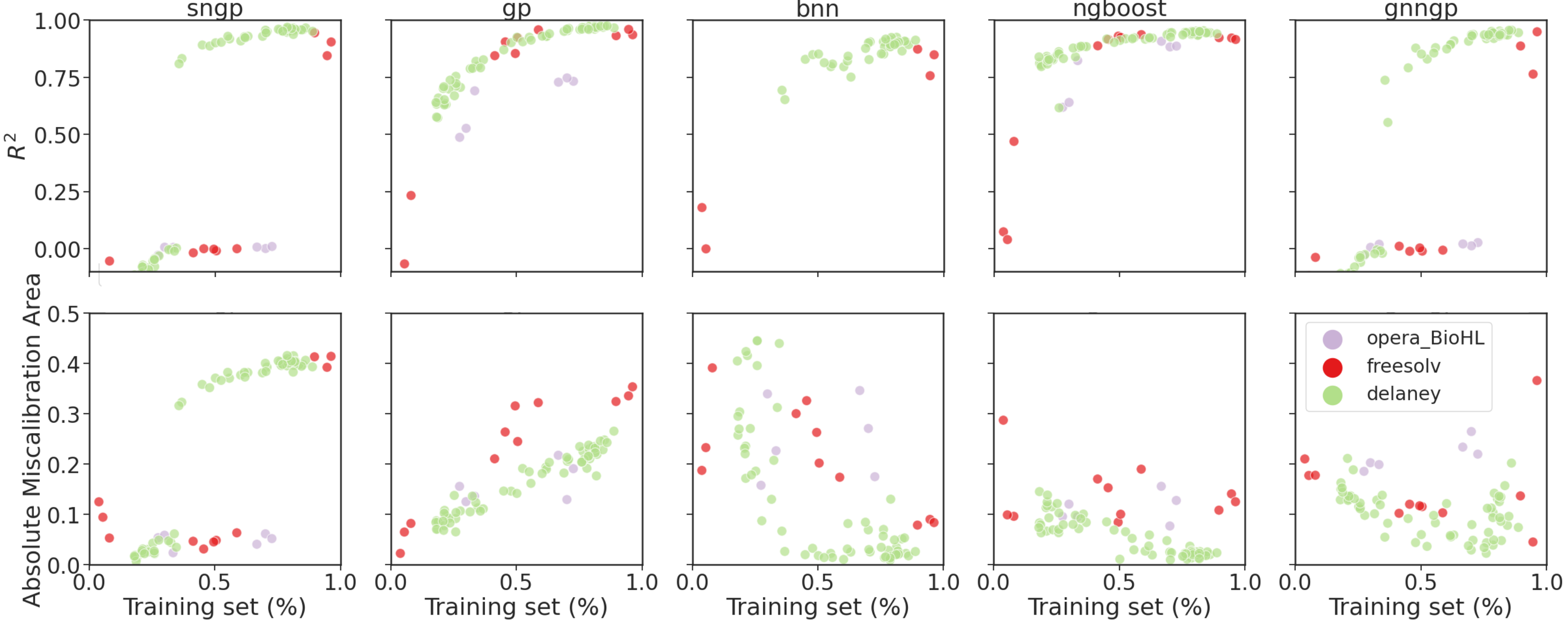}
    \mycaption{Plots of metrics of regression models on ablated cluster splits}{Graph representations are used for GNNGP, while the remaining models used Mordred descriptors.}
    \label{fig:generalization_regression}
\end{figure*}


The results of the optimization with varying $\beta$ on the Delaney dataset are shown in Figure~\ref{fig:delaney_ucb_scan}. The scans were performed only on GP and NGBoost models with Mordred descriptors, which were among the most promising model-feature combinations observed in the BO traces of Figure~\ref{fig:delaney_bo_trace}. In the BO traces (Figure~\ref{fig:delaney_ucb_scan}A), for both models, the most effective minimization is observed at $\beta = 0.25$, which corresponds to a $\beta$ parameter of 0.33 in the typical UCB acquisition function (Eq.~\ref{eq:conv_ucb}). At higher values of beta, performance quickly degrades in the GP model, while NGBoost remains performant until $\beta=0.75$. NGBoost is able to find the optimal molecule at all values of $\beta$ within the budget, and in general, performs better than GPs at data regimes of $< 50$ molecules. This seems to indicate that the prediction quality has more weight in determining the success of BO, and a more exploitative approach is warranted.


The performance and calibration metrics on a separate test set of the model at every batch of 5 evaluations are shown in Figure~\ref{fig:delaney_ucb_scan}B. Despite poorer optimization, the fully variance-based sampling strategy achieves better $R^2$ and AMA scores than the greedy strategy. The variance-based strategy suggests more diverse candidates at each iteration, allowing the models to train on a more diverse set of molecules, hence increasing performance on the test set. Models with better predictions and calibration do not necessarily give faster Bayesian optimization.

For both models, we observe the general trend of better predictions and uncertainties with more data. Overall, NGBoost achieves higher $R^2$ and lower AMA. We also observe a severe drop in early optimization performance at around batch 10 for the GPs at $\beta = 0$ and $0.25$, likely attributed to the inclusion of molecules that are further in feature space with increasing batches, after initial batches of similar molecules. This drop in performance is not observed in the NGBoost model, as random forest models can arbitrarily divide the feature space, rather than relying on a kernel function for feature distances. Interestingly, despite this pronounced drop in performance, BO with GPs achieved optimization trace results comparable to NGBoost. 




%% file: generalization.tex
\subsection{Generalizability} \label{subsec:generalizability}

Generalizability of predictive models is important in order to make accurate and reliable predictions on new chemical structures, especially in the low-data regime, where there is access to only a small slice of the chemical space. While models can predict and classify on molecules similar to the training set, we are often concerned with the performance and calibration when extrapolating to molecules that are OOD. Measuring the predictive performance of a model on the test set of a single random split only provides a partial view to its generalization capabilities---the biases in a single split can give an overconfident or underconfident estimate of performance.


To simulate prediction of OOD molecules, the models and featurizations are trained and tested on cluster splits of the datasets, as shown in Figure~\ref{fig:cluster_splits}. The clusters of molecules represent ``distributions" of similarly structured molecules and are aggregated in different combinations to create a series of training sets with difference sizes. 

\begin{figure*}[htb]
    \centering
    \includegraphics[width=0.8\textwidth]{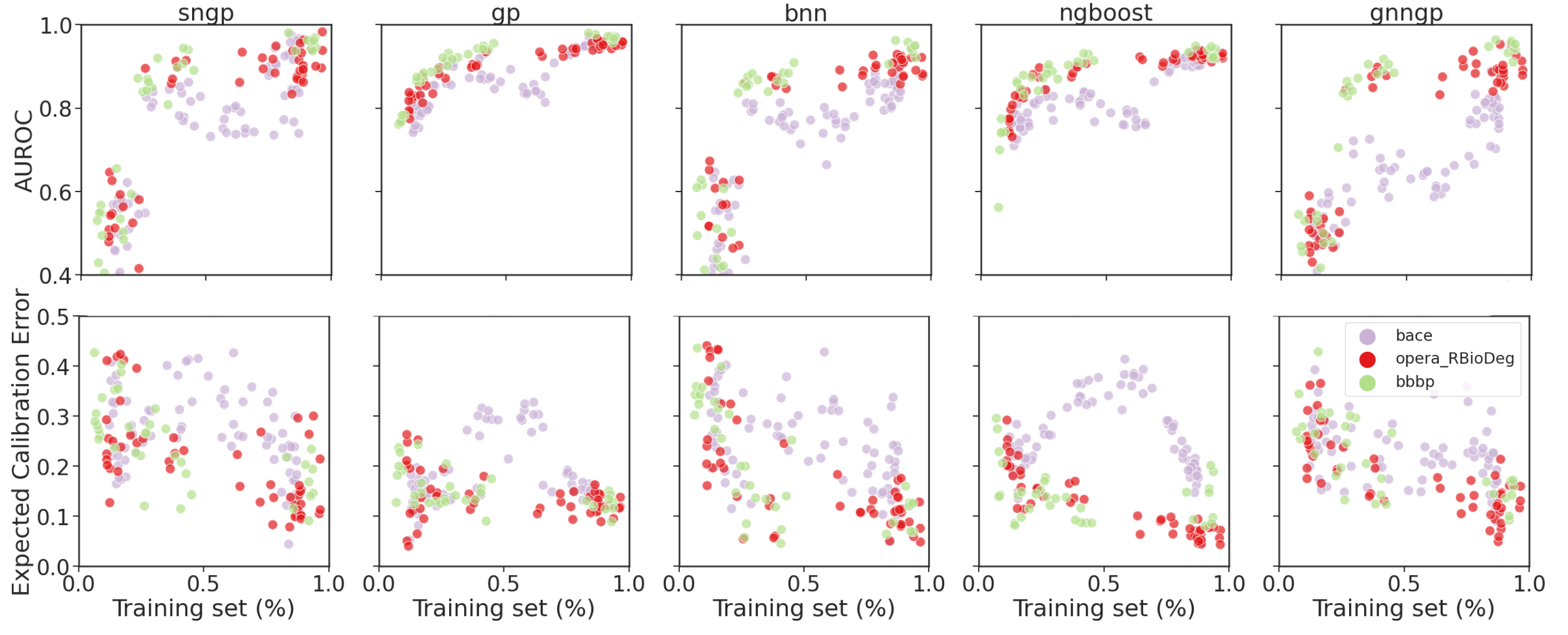}
    \mycaption{Plots of metrics of binary classification models on ablated cluster splits}{Graph representations are used for GNNGP, while the remaining models used Mordred descriptors.}
    \label{fig:generalization_classification}
\end{figure*}

Visualizations of model prediction and uncertainty quality as a function of amount of accessible training data for the regression datasets are shown in Figure ~\ref{fig:generalization_regression}. Here, only the Mordred descriptors and the graph representation (for GNNGP) are studied. As expected, we observe an increase in the $R^2$ score with increasing training data. For the smallest BioHL dataset, there are not enough clusters to form ablated sets that span the gamut of training set sizes. Deep learning models like SNGP, BNN and GNNGP are unable to achieve $R^2 > 0$ for BioHL, as previously observed in the supervised learning studies (Section ~\ref{subsec:exp_supervised_learning}). There is a clear jump in performance for deep models at around $40\%$ and $60\%$ of training set for Delaney and Freesolv, respectively. This indicates that the deep learning models require at least $\sim 300$ molecules to achieve sufficient performance. 

We observe better performance on Freesolv and BioHL using GPs and NGBoost, with NGBoost achieving higher $R^2$ and lower calibration errors, particularly when only given access to small number of clusters in the chemical space, indicating better performance at lower data regimes. However, as observed in the Delaney dataset performance, GPs are able to achieve higher $R^2$ scores once enough data is provided, indicating better generalization. In the calibration metric of the the cluster splits, we observe a general decrease in the error for BNN, and NGBoost. However, for SNGP, GNNGP, and GP, we observe an increase in AMA, particularly for the smaller datasets. A possible explanation for this: as the GP models gain more access to chemical space with more clusters, the covariance matrix determined by the kernel function gives larger uncertainties, due to low similarity of new inputs from different clusters. This results in underconfident predictions, which we observe in the reliability curves.

The generalizability results for binary classification datasets are seen in Figure~\ref{fig:generalization_classification}. As a metric of generalizability, the median performance over the cluster splits are shown in Table~\ref{tab:gen_metric}. Again, we observe that the deep learning models are only able to get decent performance at around $25-30\%$ of the the training data, corresponding to $\sim 300$ data points. The GPs and NGBoost models both achieved similar AUROC scores in this data regime, but the GPs are able to reach higher performance metrics, indicating better generalizability. For all models, there is a dip in performance for the BACE dataset, possibly due to the inclusion of clusters that are further in the feature space. This is particularly pronounced in the calibration metrics, in which there is a rise at around $50\%$ of the BACE dataset. In the calibration metrics, all models exhibit decrease in ECE with more training data, with NGBoost achieving the lowest score.

\begin{table}[!ht]
    \centering
    \begin{ruledtabular}
    \begin{tabular}{cccccc}
             & BNN & GP & NGBoost & SNGP & GNNGP \\  \hline
        BioHL & -2.511 & 0.710 & \textbf{0.853}  & 0.007 & 0.016 \\ 
        Freesolv & -0.207 & \textbf{0.916} &  0.905 & -0.001 & -0.006 \\
        Delaney & 0.828 & \textbf{0.929} &  0.920 & 0.923 & 0.880 \\
        \hline
        BACE & 0.779 & \textbf{0.874} & 0.817  & 0.780 & 0.648 \\ 
        RBioDeg & 0.876 & \textbf{0.933} &  0.909 & 0.875 & 0.873 \\
        BBBP & 0.858 & \textbf{0.921} &  0.892 & 0.856 & 0.868 \\
    \end{tabular}
    \end{ruledtabular}
    \mycaption{Metric of generalizability}{The median of performance of the models on cluster splits of each dataset using Mordred descriptors (graph representation for GNNGP). Higher value indicates better generalizability.}
    \label{tab:gen_metric}
\end{table}

%% file: conclusion.tex
\section{Conclusion} \label{sec:conclusion}

In this work, we have performed a comprehensive study of the performance and application of probabilistic models on small molecular datasets, for both regression and binary classification tasks. Several models were trained and tested on the datasets with a variety of molecular input features. We evaluate the models based on their prediction accuracy and uncertainty calibration, and their effects on a simulated experimental optimization campaign and the generalizability OOD clusters. 

Based on the results, we compile a ``handbook" of recommendations for predictive tasks with ML models on small molecular datasets:
\begin{itemize} \itemsep -3pt
    \item Mordred features are quite robust, independent of model choice. 
    \item GPs with Mordred features are a solid modelling choice for small datasets. This combination fared well in all tasks and experiments. Model setup and optimization is relatively straightforward.
    \item Out of the models tested, GPs seem to perform best on OOD molecules.
    \item NGBoost performs best for much smaller datasets ($< 100$ molecules).
    \item If using MFP features, GPs with the Tanimoto kernel provide best results.
    \item Deep learning techniques suffer from bad performance for very low data regimes ($< 300$ molecules). Their performance starts to become comparable to GPs after dataset sizes of 500 molecules. Nonetheless, these techniques require more careful setup to properly train and regularize, such as selecting training hyperparameters, and model architecture.
    \item When provided enough molecules, BNN with Mordred descriptors and GNNGP with graph tuples both give robust predictions and calibrated uncertainties.
    \item Learned graph embeddings are expressive and viable features, even at low data regimes of $\sim 150$ molecules, provided that the features are used with GPs or NGBoost.
    \item When performing Bayesian optimization, even though purely predictive models (UCB with $\beta=0$) find hits faster, their model performance is much worse than any model with some exploratory component ($\beta>0$). We found that for the UCB acquisition function on the Delaney dataset, $\beta=0.33$ tends to give best performance.
    \item Good prediction and calibration of a surrogate model on a test set does not necessarily correspond to better Bayesian optimization.
\end{itemize}

There are some caveats to our analysis that may be addressed in future work. While we only look at particular metrics for the performance and calibration, there are a number of other metrics, particularly for calibration such as negative log-likelihood or ranking coefficients between the error and the uncertainties, which may provide different perspectives for the observed results. Additionally, we do not perform any optimization of the hyperparameters or architectures, which would typically be done for each model, dataset, and molecule representation. For other future work, besides the addition of more models and features, the study can be extended to multi-classification molecular tasks. Regardless of these potential future extensions, we believe that the work here presented here provides important insights to the development and application of probabilistic models on low data chemical datasets.
%




%% file: si.tex
\section{Supplementary information}

\subsection{Graph features}

\begin{table}[htb] 
\centering
\begin{tabular}{ l @{\qquad} l }
\toprule
Node features &  Categories   \\ 
\hline 
Atomic number  &  one-hot encoding from set of heavy atoms in dataset  \\
Chirality  & unspecified, CW, CCW, \emph{UNK} \\
Atom degree & 0, \dots, 10, \emph{UNK} \\
Formal charge & -5, \dots, 5, \emph{UNK}    \\
Number of hydrogens & 0, \dots, 8, \emph{UNK}  \\
Number of radical electrons & 0, \dots, 4, \emph{UNK} \\
Hybridization & sp, sp2, sp3, sp3d, sp3d2, \emph{UNK} \\
Is aromatic & True/False\\
Part of ring & True/False \\
\hline
\hline
Edge features & Categories \\
\hline
Bond type  &  single, double, triple, aromatic, \emph{UNK}\\
Stereo configuration  & none, $Z$, $E$, \emph{cis}, \emph{trans}, any \\
Is conjugated & True/False \\
\botrule
\end{tabular}
\caption{Features for the vertex and edge features of graph tuple. All cateogries are one-hot encoded and stacked to give a singular bit vector. \emph{UNK} stands for ``unknown", and is a catch-all category.}
\label{tab:node-edge-features}
\end{table}


\clearpage

\subsection{Performance and Calibration Metrics}

\begin{table}[h!] 
 \begin{center}
\begin{tabular}{ c @{\qquad} c @{\qquad} c @{\qquad} c } \toprule
\textbf{BioHL} & MFP & Mordred & Graph-based   \\ 
\hline 
SNGP & $-0.107^{+0.114}_{-0.432}$  & $-0.136^{+0.021}_{-0.466}$ & $-0.136^{+0.092}_{-0.437}$  \\
GP & $0.383^{+0.308}_{-0.557}$   & $0.817^{+0.131}_{-0.242}$  &  $0.750^{+0.160}_{-0.279}$  \\
BNN & $0.011^{+0.197}_{-0.436}$  & $-0.103^{+0.594}_{-1.378}$ & $-0.066^{+0.296}_{-0.818}$   \\
NGBoost & $0.320^{+0.373}_{-0.786}$ & $0.843^{+0.096}_{-0.181}$ & $0.803^{+0.142}_{-0.228}$    \\
GNNGP & \done & \done & $-0.129^{+0.146}_{-0.458}$  \\
\hline
\textbf{Freesolv} & MFP & Mordred & Graph-based  \\ 
\hline 
SNGP & $0.738^{+0.093}_{-0.081}$  & $0.912^{+0.029}_{-0.026}$ & $0.891^{+0.056}_{-0.039}$   \\
GP & $0.716^{+0.130}_{-0.138}$   & $0.924^{+0.031}_{-0.050}$  &  $0.875^{+0.039}_{-0.051}$  \\
BNN & $0.601^{+0.097}_{-0.086}$  & $0.673^{+0.104}_{-0.110}$ & $0.845^{+0.044}_{-0.071}$   \\
NGBoost & $0.556^{+0.131}_{-0.128}$ & $0.925^{+0.027}_{-0.044}$ & $0.887^{+0.038}_{-0.053}$    \\
GNNGP & \done & \done  & $0.903^{+0.039}_{-0.039}$  \\
\hline
\textbf{Delaney} & MFP & Mordred & Graph-based  \\ 
\hline
SNGP & $0.687^{+0.066}_{-0.073}$  & $0.918^{+0.021}_{-0.025}$ & $0.904^{+0.023}_{-0.027}$   \\
GP & $0.724^{+0.052}_{-0.053}$   & $0.934^{+0.016}_{-0.018}$  &  $0.897^{+0.025}_{-0.030}$  \\
BNN & $0.687^{+0.055}_{-0.061}$  & $0.908^{+0.019}_{-0.023}$ & $0.905^{+0.029}_{-0.030}$   \\
NGBoost & $0.486^{+0.070}_{-0.076}$ & $0.915^{+0.019}_{-0.025}$ & $0.897^{+0.029}_{-0.030}$   \\
GNNGP &\done & \done & $0.911^{+0.027}_{-0.022}$  \\
\botrule
\end{tabular}
\qquad 
\begin{tabular}{ c @{\qquad} c @{\qquad} c @{\qquad} c } \toprule
\textbf{BioHL} & MFP & Mordred & Graph-based   \\ 
\hline 
SNGP & $0.070^{+0.052}_{-0.031}$  & $0.071^{+0.055}_{-0.032}$ & $0.075^{+0.049}_{-0.035}$  \\
GP & $0.061^{+0.060}_{-0.033}$   & $0.141^{+0.098}_{-0.074}$  &  $0.200^{+0.097}_{-0.098}$  \\
BNN & $0.363^{+0.069}_{-0.089}$  & $0.103^{+0.088}_{-0.065}$ & $0.217^{+0.088}_{-0.096}$   \\
NGBoost & $0.109^{+0.098}_{-0.058}$ & $0.109^{+0.057}_{-0.046}$ & $0.086^{+0.088}_{-0.053}$    \\
GNNGP &\done &\done & $0.211^{+0.101}_{-0.101}$  \\
\hline
\textbf{Freesolv} & MFP & Mordred & Graph-based  \\ 
\hline 
SNGP & $0.628^{+0.033}_{-0.033}$  & $0.359^{+0.024}_{-0.026}$ & $0.345^{+0.028}_{-0.031}$   \\
GP & $0.147^{+0.040}_{-0.046}$   & $0.106^{+0.048}_{-0.048}$  &  $0.271^{+0.035}_{-0.037}$  \\
BNN & $0.243^{+0.051}_{-0.051}$  & $0.112^{+0.040}_{-0.037}$ & $0.147^{+0.039}_{-0.038}$   \\
NGBoost & $0.032^{+0.026}_{-0.016}$ & $0.052^{+0.028}_{-0.023}$ & $0.035^{+0.031}_{-0.018}$    \\
GNNGP & \done& \done& $0.086^{+0.047}_{-0.044}$  \\
\hline
\textbf{Delaney} & MFP & Mordred & Graph-based  \\ 
\hline
SNGP & $0.202^{+0.029}_{-0.030}$  & $0.347^{+0.017}_{-0.019}$ & $0.326^{+0.018}_{-0.020}$   \\
GP & $0.080^{+0.037}_{-0.032}$   & $0.120^{+0.032}_{-0.036}$  &  $0.110^{+0.038}_{-0.038}$  \\
BNN & $0.223^{+0.042}_{-0.039}$  & $0.166^{+0.036}_{-0.040}$ & $0.044^{+0.022}_{-0.019}$   \\
NGBoost & $0.070^{+0.038}_{-0.036}$ & $0.031^{+0.032}_{-0.017}$ & $0.056^{+0.036}_{-0.028}$   \\
GNNGP & \done& \done& $0.055^{+0.020}_{-0.019}$   \\
\botrule
\end{tabular}
\mycaption{Performance and calibration results on regression datasets}{(\emph{left}) $R^2$ metric and (\emph{right}) AMA for each feature and model pair. Graph tuples were used for GNNGP, while graph embeddings were used for all other models. The 95\% confidence interval is reported.}
\label{tab:regression_performance}
\end{center}
\end{table}

\begin{table}[h!] 
 \begin{center} 
\begin{tabular}{ c @{\qquad} c @{\qquad} c @{\qquad} c } \toprule
\textbf{BACE} & MFP & Mordred & Graph-based   \\ 
\hline 
SNGP & $0.890_{-0.037}^{+0.036}$  & $0.879_{-0.041}^{+0.039}$ & $0.873_{-0.040}^{+0.039}$  \\
GP & $0.917_{-0.031}^{+0.029}$  & $0.915_{-0.032}^{+0.028}$   &  $0.865_{-0.041}^{+0.041}$  \\
BNN & $0.918_{-0.029}^{+0.028}$   & $0.893_{-0.037}^{+0.034}$ & $0.869_{-0.041}^{+0.040}$    \\
NGBoost & $0.895_{-0.034}^{+0.031}$  & $0.890_{-0.039}^{+0.032}$  & $0.857_{-0.046}^{+0.039}$   \\
GNNGP &\done & \done& $0.845_{-0.046}^{+0.041}$   \\
\hline
\textbf{RBioDeg} & MFP & Mordred & Graph-based  \\ 
\hline 
SNGP & $0.783 _{-0.055}^{+0.053}$  & $0.832 _{-0.048}^{+0.042}$ & $0.826 _{-0.048}^{+0.042}$  \\
GP & $0.837 _{-0.048}^{+0.040}$  & $0.858 _{-0.042}^{+0.039}$   &  $0.835 _{-0.044}^{+0.043}$  \\
BNN & $0.826 _{-0.045}^{+0.042}$   & $0.852 _{-0.043}^{+0.038}$ & $0.833 _{-0.044}^{+0.040}$  \\
NGBoost & $0.791 _{-0.053}^{+0.050}$  & $0.846 _{-0.042}^{+0.042}$  & $0.829 _{-0.049}^{+0.046}$  \\
GNNGP &\done &\done & $0.840 _{-0.048}^{+0.044}$  \\
\hline
\textbf{BBBP} & MFP & Mordred & Graph-based  \\ 
\hline
SNGP & $0.882 _{-0.050}^{+0.041}$  & $0.902 _{-0.039}^{+0.036}$ & $0.870 _{-0.053}^{+0.047}$  \\
GP & $0.910 _{-0.040}^{+0.035}$  & $0.922 _{-0.037}^{+0.0.35}$   &  $0.894 _{-0.045}^{+0.038}$  \\
BNN & $0.886 _{-0.046}^{+0.043}$   & $0.900 _{-0.049}^{+0.043}$ & $0.883 _{-0.050}^{+0.043}$  \\
NGBoost & $0.842 _{-0.053}^{+0.051}$  & $0.896 _{-0.048}^{+0.042}$  & $0.834 _{-0.053}^{+0.050}$  \\
GNNGP & \done& \done& $0.879 _{-0.051}^{+0.043}$  \\
\botrule
\end{tabular}
\qquad 
\begin{tabular}{ c @{\qquad} c @{\qquad} c @{\qquad} c } \toprule
\textbf{BACE} & MFP & Mordred & Graph-based   \\ 
\hline 
SNGP & $0.184_{-0.070}^{+0.075}$  & $0.191_{-0.065}^{+0.068}$ & $0.155_{-0.043}^{+0.051}$  \\
GP & $0.093_{-0.032}^{+0.039}$  & $0.090_{-0.035}^{+0.041}$   &  $0.155_{-0.052}^{+0.062}$  \\
BNN & $0.141_{-0.049}^{+0.059}$   & $0.171_{-0.040}^{+0.040}$ & $0.146_{-0.056}^{+0.060}$    \\
NGBoost & $0.095_{-0.033}^{+0.038}$  & $0.076_{-0.032}^{+0.037}$  & $0.171_{-0.071}^{+0.089}$   \\
GNNGP & \done & \done & $0.127_{-0.045}^{+0.050}$   \\
\hline
\textbf{RBioDeg} & MFP & Mordred & Graph-based  \\ 
\hline 
SNGP & $0.147_{-0.048}^{+0.056}$  & $0.249_{-0.074}^{+0.078}$ & $0.244_{-0.080}^{+0.086}$  \\
GP & $0.075_{-0.030}^{+0.037}$  & $0.096_{-0.038}^{+0.044}$   &  $0.132_{-0.046}^{+0.051}$  \\
BNN & $0.104_{-0.040}^{+0.043}$   & $0.097_{-0.039}^{+0.046}$ & $0.142_{-0.050}^{+0.052}$    \\
NGBoost & $0.081_{-0.034}^{+0.038}$  & $0.150_{-0.055}^{+0.055}$  & $0.125_{-0.048}^{+0.053}$   \\
GNNGP & \done & \done & $0.223_{-0.058}^{+0.063}$   \\
\hline
\textbf{BBBP} & MFP & Mordred & Graph-based  \\ 
\hline
SNGP & $0.160 _{-0.059}^{+0.064}$  & $0.263 _{-0.074}^{+0.086}$ & $0.290 _{-0.115}^{+0.119}$  \\
GP & $0.127 _{-0.046}^{+0.061}$  & $0.112 _{-0.043}^{+0.051}$   &  $0.165 _{-0.026}^{+0.027}$  \\
BNN & $0.230 _{-0.095}^{+0.108}$  & $0.118 _{-0.053}^{+0.067}$  & $0.282 _{-0.100}^{+0.110}$  \\
NGBoost & $0.168 _{-0.058}^{+0.059}$   & $0.188 _{-0.083}^{+0.089}$ & $0.368 _{-0.150}^{+0.118}$  \\
GNNGP & \done & \done & $0.246 _{-0.063}^{+0.073}$   \\
\botrule
\end{tabular}
\mycaption{Performance and calibration results on binary classification datasets}{(\emph{left}) AUROC metric and (\emph{right}) ECE for each feature and model pair. Graph tuples were used for GNNGP, while graph embeddings were used for all other models. The 95\% confidence interval is reported.}
\label{tab:binary_performance}
\end{center}
\end{table}

\clearpage

\subsection{Bayesian Optimization Traces}
\label{sisubsec:bo_traces}

\begin{figure}[h!]
\begin{subfigure}[b]{0.35\textwidth}
    \centering
    \includegraphics[width=\textwidth]{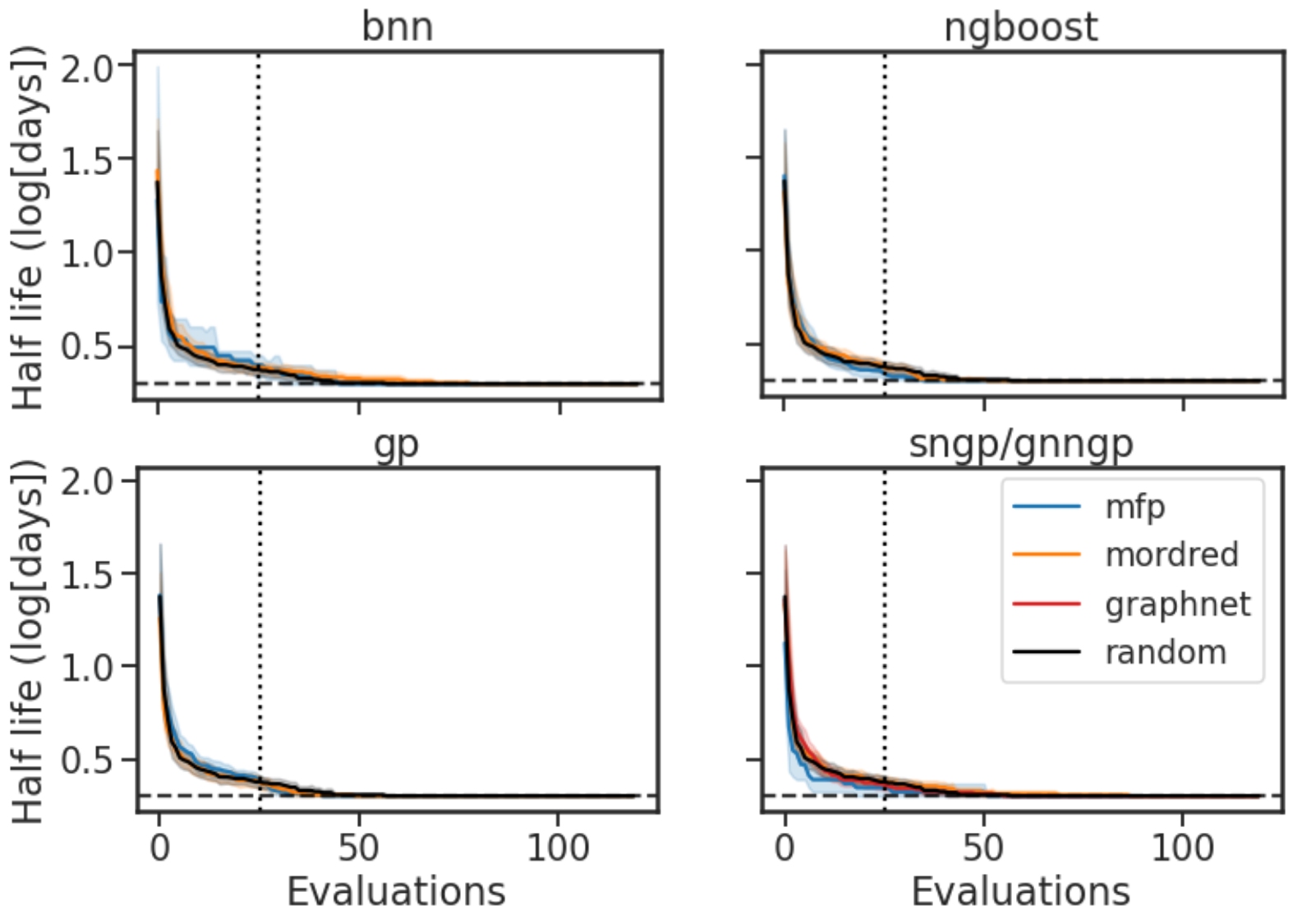}
    \caption{BioHL dataset. Minimizing the log half-life of biodegredation.}
\end{subfigure}
\qquad
\begin{subfigure}[b]{0.37\textwidth}
    \centering
    \includegraphics[width=\textwidth]{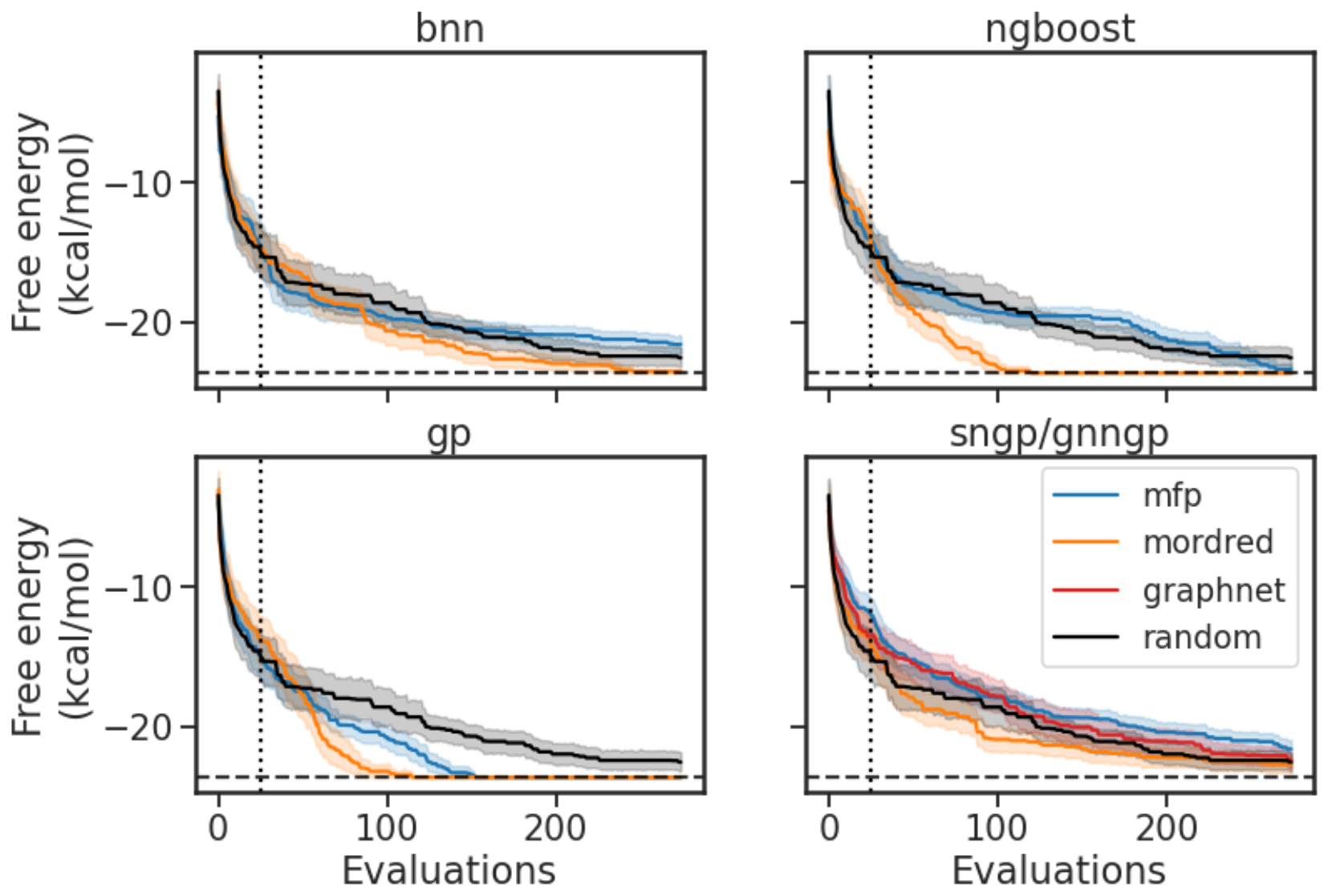}
    \caption{Freesolv dataset. The goal is to minimize the free energy of solvation.}
\end{subfigure}
\mycaption{BO traces for regression datasets}{Traces show average values over 30 independent runs, and shaded area is the 95\% confidence interval. The BO experiments start with randomly sampled 5\% of the dataset (minimum of 25 molecules), represented by the vertical dashed lines. The optimal molecule half-life is shown by the horizontal dashed lines.}
\end{figure}

\begin{figure}[h!]
\begin{subfigure}[b]{0.32\textwidth}
    \centering
    \includegraphics[width=\textwidth]{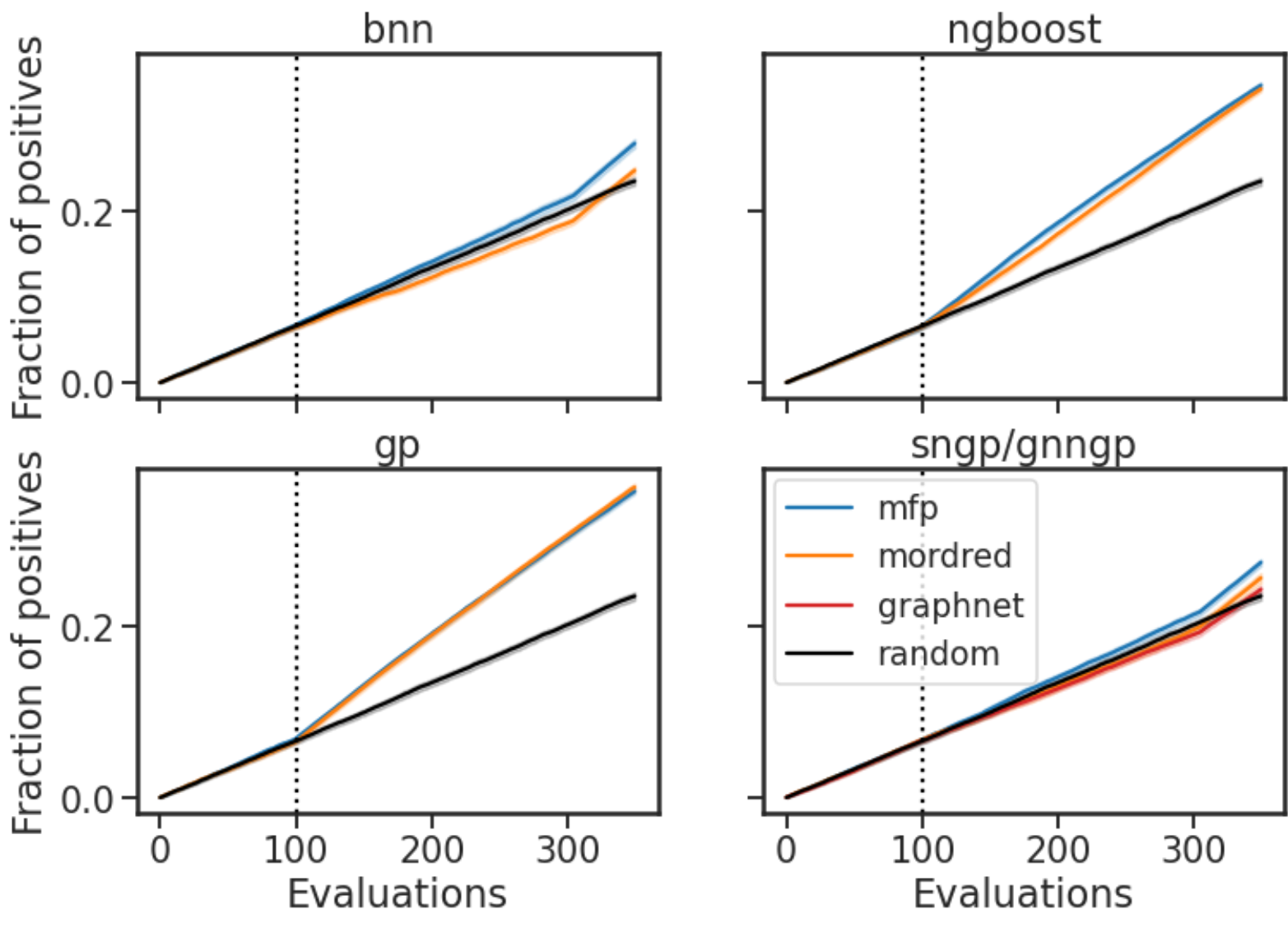}
    \caption{BACE dataset. Find as many proteins that dock to BACE-1.}
\end{subfigure}
\hfill
\begin{subfigure}[b]{0.32\textwidth}
    \centering
    \includegraphics[width=\textwidth]{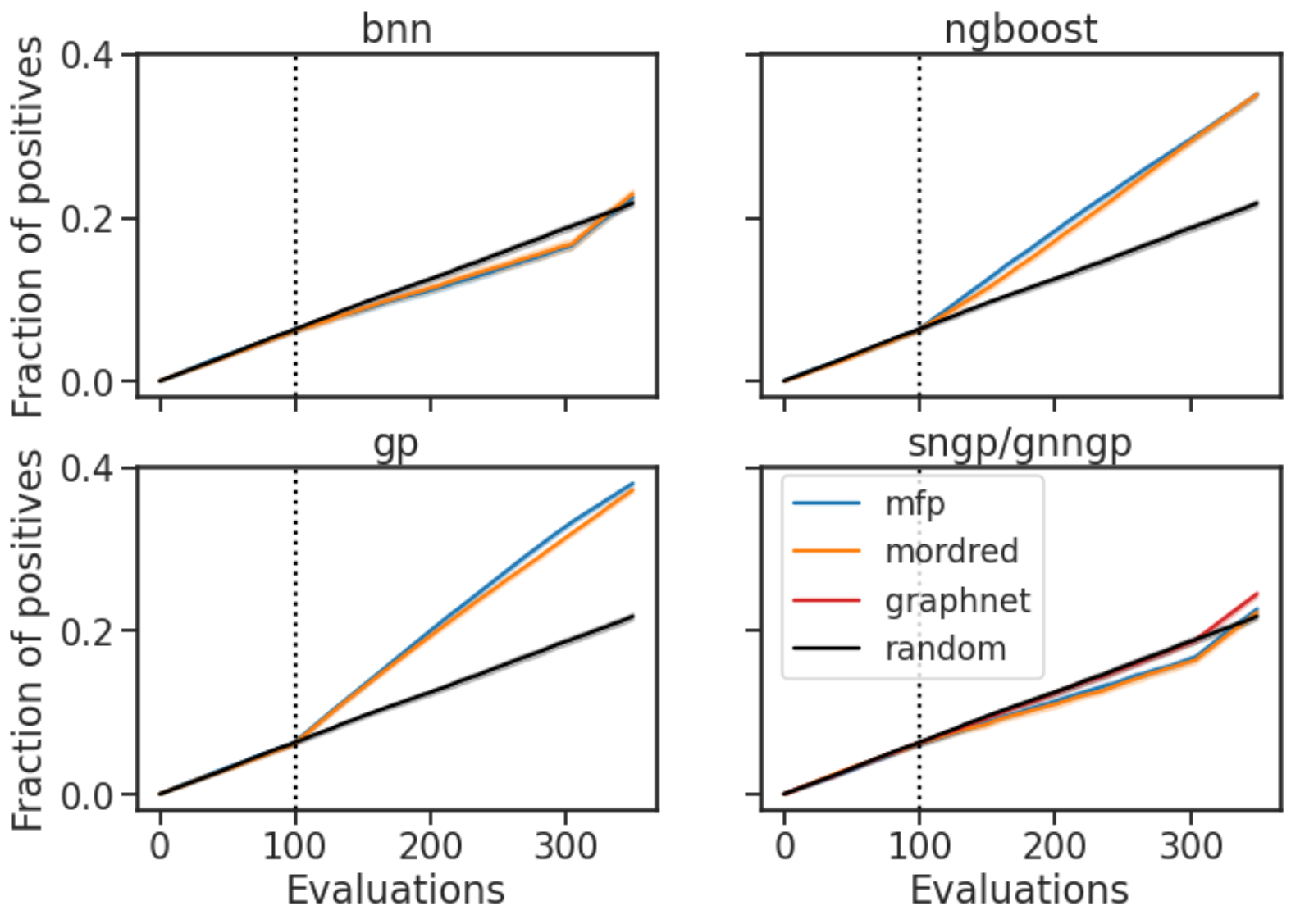}
    \caption{RBioDeg dataset. Find as many molecules that biodegrades.}
\end{subfigure}
\hfill
\begin{subfigure}[b]{0.32\textwidth}
    \centering
    \includegraphics[width=\textwidth]{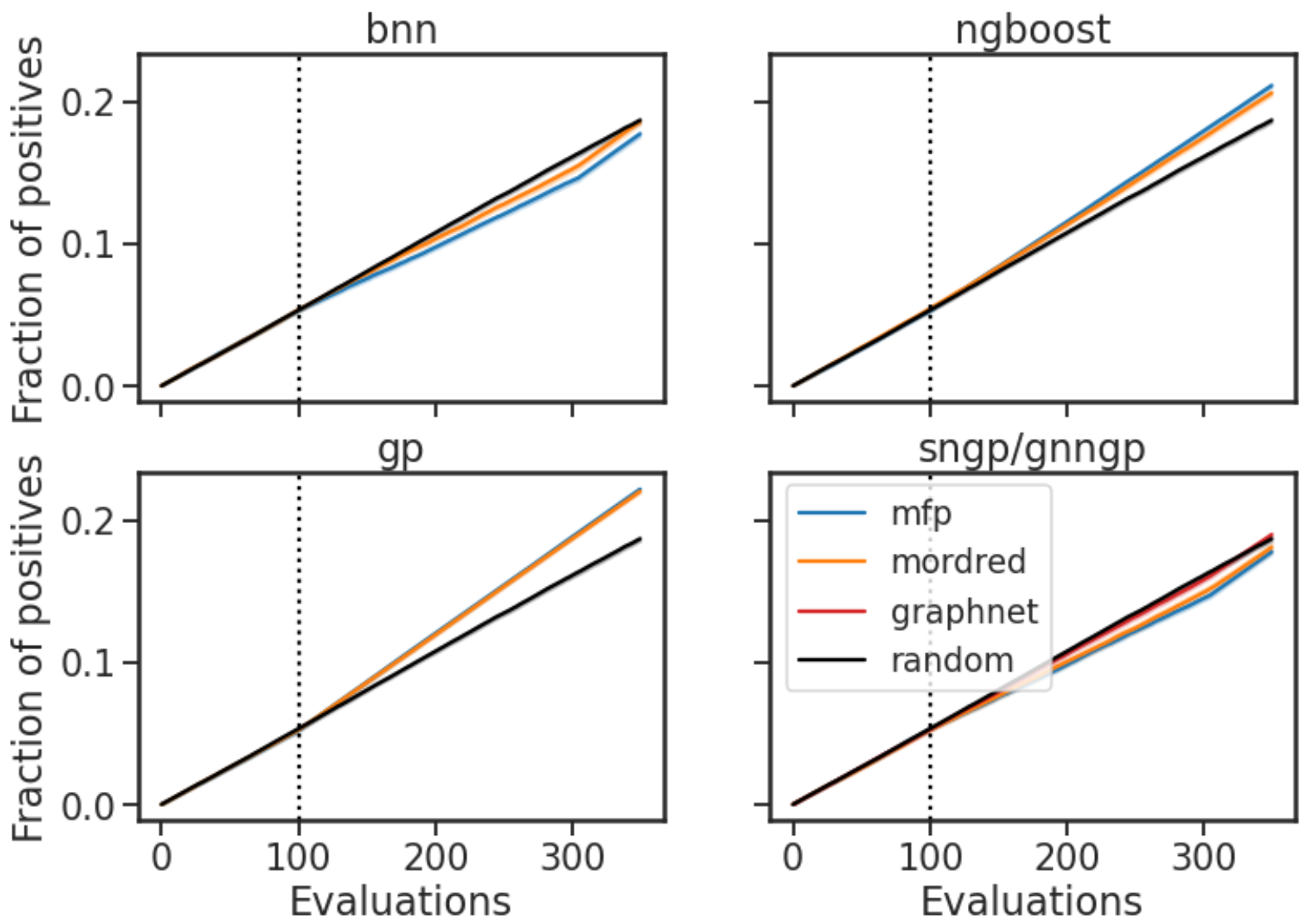}
    \caption{BBBP dataset. Find the molecules that can cross blood-brain barrier.}
\end{subfigure}
\mycaption{BO traces for binary classification datasets}{Traces show average values over 30 independent runs, and shaded area is the 95\% confidence interval. The BO experiments start with randomly sampled 10\% of the dataset (maximum of 100 molecules), represented by the vertical dashed lines.}
\end{figure}

\clearpage


\subsection{Generalizability} \label{sisubsec:generalizability}



\begin{figure}[h!]
    \centering
    \includegraphics[width=0.5\textwidth]{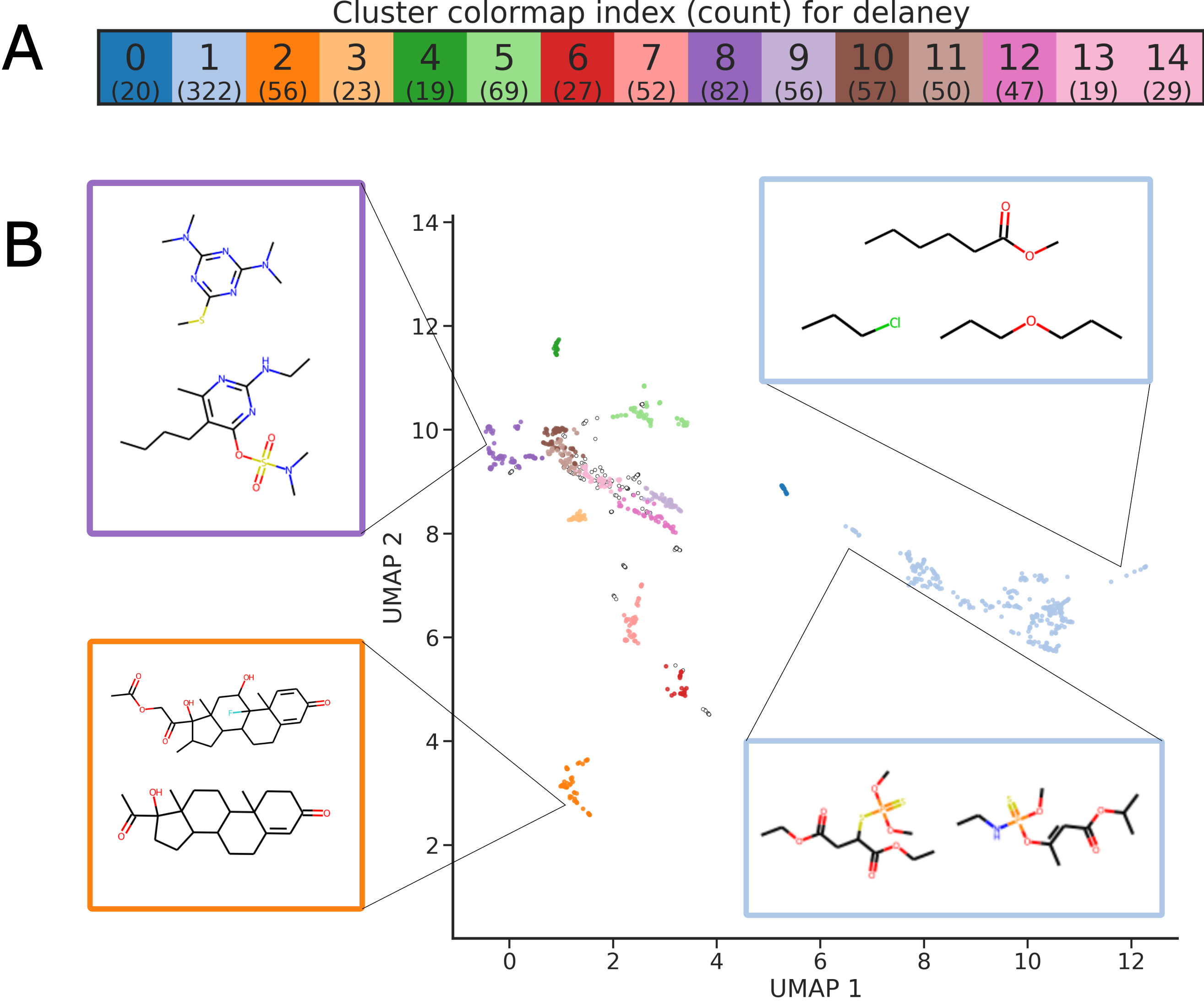}
    \mycaption{Cluster generated for cluster splits on the Delaney dataset}{A) Clusters identified by HDBScan algorithm,\cite{campello2013density} coloured and labelled, with the number of molecules per cluster listed. B) Visualization of UMAP~\cite{mcinnes_umap_2018,mcinnes_umap_2020} reduced chemical space, with samples of molecules from clusters shown. Similar clusters have similar structures. Molecules further in chemical space have more structural differences.}
\end{figure}